\documentclass[10pt,letterpaper,twocolumn]{article}
\usepackage[T1]{fontenc}
\usepackage{lmodern}
\usepackage[hmargin=0.69in,top=0.72in,bottom=0.76in,columnsep=0.28in]{geometry}
\usepackage{microtype}
\usepackage{amsmath,amssymb,graphicx}
\usepackage{cite}
\usepackage[hidelinks]{hyperref}
\newcommand{\papertitle}{Discrete event-chain Monte Carlo algorithm}
\hypersetup{pdftitle={\papertitle},pdfauthor={Hidemaro Suwa}}

\newcommand{\av}[1]{\left\langle #1\right\rangle}
\newcommand{\tint}{\tau_{\mathrm{int}}}
\newcommand{\doi}[2]{\href{https://doi.org/#1}{#2}}
\title{\Large\bfseries\papertitle}
\author{Hidemaro Suwa\\[3pt]
\small Department of Physics, The University of Tokyo, Tokyo 113-0033, Japan\\
\small \href{mailto:suwamaro@phys.s.u-tokyo.ac.jp}{suwamaro@phys.s.u-tokyo.ac.jp}}
\date{}
\begin{document}
\twocolumn[{
\begin{@twocolumnfalse}
\maketitle

\begin{abstract}
Event-chain Monte Carlo methods keep an active variable and a direction of motion across successive updates, but their constructions for particles and continuous spins rely on infinitesimal moves. We construct an event-chain algorithm for interacting discrete variables. The active variable changes by one step along an ordered set of states, activity is handed to a neighbor at a rate set by a prescribed bond current, and the direction reverses only to absorb the remaining rate imbalance, the residual. All rates follow from single-site conditional probabilities, and centering the bond current over the conditional law minimizes the mean reversal rate. Cumulative matching of residuals of opposite sign along a site order further turns reversals into longer-range handoffs. We compare the integrated autocorrelation time of the longest-wavelength structure factor with those of Metropolis, heat bath, irreversible Metropolized Gibbs, and Suwa--Todo samplers. The event chain lowers the size exponent of this time from about 2 to $0.503\pm0.016$ in the one-dimensional discrete Gaussian (DG) model, where the direction never reverses, and to approximately 1.5 in the cubic-lattice six-state clock model. In the cubic-lattice DG model the time stays below half an event sweep and does not grow with system size, whereas those of the random-site samplers grow with exponents of 1.6--2.0. Where the exponent is unchanged, the event chain shortens the time measured in elementary events and raises the number of effective samples per core-second by factors of 43--89 and 58--72 near the roughening transition of the square-lattice DG model, 5.2--11 and 2.4--3.6 in the square-lattice clock model, 11--25 and 3.5--5.2 at the square-lattice Ising critical point, and 3.2--5.7 and 1.3--2.1 at the three-state Potts critical point.
\end{abstract}
\vspace{0.7\baselineskip}
\end{@twocolumnfalse}
}]

\section{Introduction}
Markov chain Monte Carlo (MCMC) methods are the standard tool for computing equilibrium averages of many interacting degrees of freedom, from lattice spin models and fluctuating interfaces to adsorbed layers and molecular charge states \cite{metropolis,hastings,sokal}. Their efficiency is limited by the slowest modes. Long-wavelength fluctuations relax slowly under local updates: near the roughening transition of an interface, for example, a height fluctuation whose wavelength spans the surface requires many local height changes \cite{chui,rough,rough2}. When the sign of each change is chosen independently, raising a site and later lowering it restores the same profile, and the mode evolves diffusively. Similar cancellations slow the sampling of long-wavelength order-parameter fluctuations near a continuous ordering transition. The number of effectively independent measurements per unit computing time depends on how many updates are needed to decorrelate the observable and on how much each update costs. Sampling efficiency is therefore characterized both by the integrated autocorrelation time $\tint$ in sweeps \cite{hohenberg,sokal} and by the number of effective samples per unit computing time.

In continuous fields, overrelaxation updates suppress this diffusive behavior and can carry a disturbance by a sequence of local moves \cite{creutz,brown}. A more general strategy, which applies also to discrete variables, is to give up detailed balance while keeping global balance. Lifting enlarges the state space with an auxiliary direction, so that a chain persists in one direction instead of backtracking \cite{chen,diaconis,gustafson}. Skewed detailed balance realizes lifting for Ising updates \cite{turitsyn,fernandes,sakaiising,curieweiss}, and geometric allocation reduces rejections by constructing transition probabilities that violate detailed balance \cite{suwa,allocation}. How often the direction changes controls the gain in relaxation and in estimator variance \cite{sakairing,gagnon,vucelja,flowreview}.

Event-chain Monte Carlo methods carry this idea from the state space into real space. In particle systems and continuous spin models, a single active variable moves continuously in a fixed direction until an interaction would oppose the motion; the opposing interaction then passes the activity, together with its direction, to its partner \cite{bernard,michel,krauthreview,spin}. A disturbance can thus travel through the system while every local change keeps the same sign, and the interaction itself decides where the next update takes place. The same mechanism applies to real-valued lattice fields, where a bond veto transfers the activity to the other endpoint \cite{realfield}. These constructions rely on infinitesimal moves, so that at any instant at most one interaction opposes the motion. Lifted exclusion processes transfer activity at finite spacing on a one-dimensional particle lattice \cite{essler,localtasep,softtasep}, and many-particle event chains allocate transfers among the participants of one interaction \cite{harland}.

For discrete variables, irreversible samplers have developed along a different line. Heat bath draws the updated variable from its single-site conditional distribution, the Suwa--Todo algorithm reduces the rejection rate for the same conditional weights \cite{suwa}, and the irreversible Metropolized Gibbs sampler (IMGS) of Faizi et al.\ keeps a direction of change across updates \cite{faizi}. Power and Vorstrup Goldman's discrete coordinate sampler can move one coordinate repeatedly and uses rate imbalances to select a generator at a switch \cite{power}. Neighborhood-weighted proposals \cite{zanella,grathwohl} and kinetic or momentum-based algorithms \cite{kinetic,hamiltonian,dhams} guide finite moves in other ways. Nonlocal updates follow yet another route: cluster algorithms rotate or reflect an extended region \cite{swendsen,wolffcluster,evertz}, and worm algorithms accumulate changes along the path of a defect \cite{classicalworm}. In the single-site samplers, however, the next site to update is chosen at random or in a fixed sweep order, independently of the change just made. A persistent direction of change is therefore not passed on to the neighbors that the change affects, as it is in an event chain, where the interaction decides the next site. Carrying over the event-chain mechanism is not straightforward. A finite step of a discrete variable can raise the energy of several bonds at once, so there is no unique interaction that vetoes the move and receives the activity.

\begin{figure*}[!t]
\centering\includegraphics[width=\linewidth]{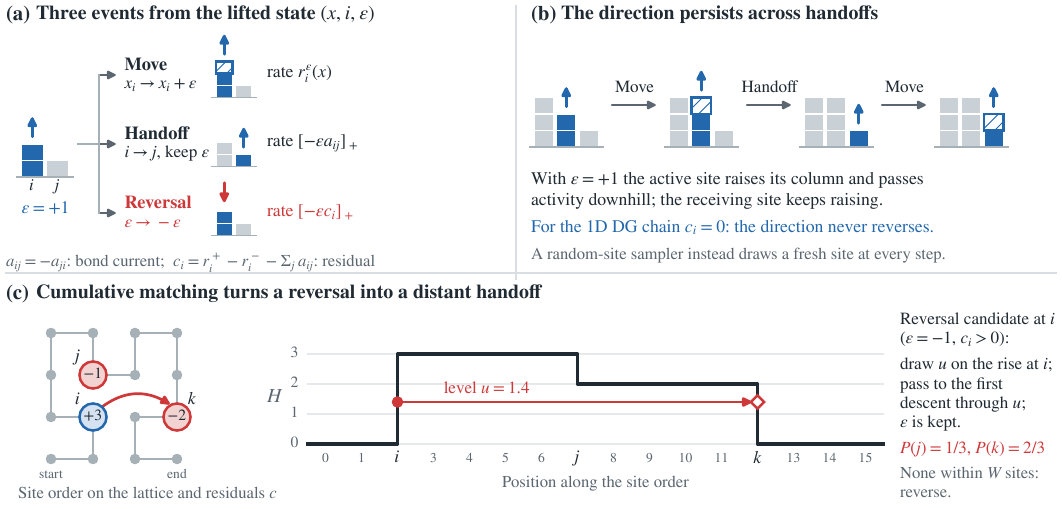}
\caption{Discrete event-chain moves, illustrated with integer heights drawn as columns. The active site is blue, and its arrow gives the direction $\epsilon$. (a) Three events compete in the lifted state $(x,i,\epsilon)$: a unit move of $x_i$ in direction $\epsilon$, a handoff that passes the activity to a neighbor $j$ and keeps $\epsilon$, and a reversal of $\epsilon$. The bond current $a_{ij}$ sets the handoff rate and the residual $c_i$ sets the reversal rate, where $[u]_+=\max(0,u)$ (Sec.~\ref{sec:models}). For the DG model in the positive direction, activity passes from a higher to a lower column. (b) Because a handoff keeps $\epsilon$, successive moves at different sites have the same sign, and the profile changes coherently. (c) Cumulative matching (Sec.~\ref{sec:pair}). Sites are visited in a fixed cyclic order, here a Hilbert curve on a periodic $4\times4$ lattice whose end is adjacent to its start through the boundary, and $H$ is the running sum of the residuals along that order. At a reversal candidate at $i$ with $\epsilon=-1$ and $c_i=+3$, a real level $u$ is drawn uniformly on the rise of $H$ at $i$, here $0<u<3$, and the activity is handed to the first site at which $H$ descends through $u$, with $\epsilon$ unchanged. The level $u=1.4$ selects $k$; if the search window reaches $k$, $j$ is selected with probability $1/3$ and $k$ with probability $2/3$. If no recipient lies within the search window of $W$ sites, the direction reverses.}\label{fig:transfer}
\end{figure*}

Here we construct an event-chain algorithm for interacting discrete variables that resolves this difficulty by working with rates rather than with vetoes of individual bonds (Fig.~\ref{fig:transfer}). The active variable moves by one step along an ordered set of its states, a prescribed antisymmetric current on each bond hands the activity and its direction to a neighbor, and the direction reverses only to absorb the remaining rate imbalance, which we call the residual. All rates follow from the single-site conditional distribution, which also fixes the minimum mean reversal rate. In the one-dimensional discrete Gaussian (DG) model this minimum is zero, so the direction never reverses. Where residuals remain, \emph{cumulative matching} replaces part of the reversals by longer-range handoffs. Both samplers preserve the target distribution.

We test the construction on interface fluctuations in the DG model, orientational fluctuations in a six-state clock model, and long-wavelength spin fluctuations in Ising and three-state Potts models, whose local states form an integer line, a cycle, or a finite sequence with reflecting ends. The interface model describes surface roughening \cite{chui}, and the Ising model also describes a lattice gas of adsorbed particles \cite{leeyang}. We compare each event chain with Metropolis, heat bath, IMGS, and the Suwa--Todo algorithm. The event chains lower the size exponent of the relaxation time of long-wavelength fluctuations in the one-dimensional DG model, and the effective exponent over the simulated sizes in the cubic-lattice clock model. In the rough phase of the square-lattice DG model and in the cubic-lattice DG model, they extend the range of sizes over which the relaxation time grows only weakly with size. Where the exponent is not changed appreciably over the simulated sizes, as near roughening and in the square-lattice clock, Ising, and Potts models, they shorten the relaxation time measured in elementary events and raise the number of effective samples per core-second by substantial factors. Comparing the two event chains at identical move rates isolates the cost and benefit of matching.

Section~\ref{sec:models} constructs the nearest-neighbor event chain and derives the minimum reversal rate. Section~\ref{sec:pair} introduces cumulative matching. Section~\ref{sec:methods} describes the reference samplers and the statistical analysis. The following two sections present the benchmarks and discuss them. The appendices prove global balance and give implementation and statistical details.

\section{Nearest-neighbor lifting for finite steps}\label{sec:models}
\subsection{Lifted state and the three events}

Let $x=(x_1,\ldots,x_N)$ be a configuration of discrete variables with target probability $\pi(x)$. We call a coordinate index a \emph{site} and specify which pairs of sites, the \emph{bonds}, can exchange activity. The sampler augments $x$ with an active site $i$ (also called the pointer) and a direction $\epsilon=\pm1$, and targets the lifted distribution $\pi(x)/(2N)$, in which $i$ and $\epsilon$ are uniform and independent of $x$. We refer to the states with a given $\epsilon$ as the direction sector $\epsilon$.

A move changes $x_i$ by one link in its chosen label order and keeps $i$ and $\epsilon$ fixed. Let $k_i^\epsilon(x,x')$ be the rate of such a move. On each physical link we impose
\begin{equation}
 \pi(x)k_i^\epsilon(x,x')=\pi(x')k_i^{-\epsilon}(x',x). \label{eq:link}
\end{equation}
Equation~\eqref{eq:link} is the continuous-time \emph{skewed detailed balance} condition \cite{turitsyn,faizi}: the equilibrium flux from $x$ to $x'$ in one direction sector equals that of the return step in the opposite sector. A move follows the label order in sector $+$ and retraces it in sector $-$. On a cycle the order wraps around, while a move directed out of a finite line is an outward attempt that reflects at the endpoint.

Moves alone do not conserve probability within a sector. Let $r_i^\epsilon(x)$ be the total move attempt rate at site $i$ in sector $\epsilon$, including an outward attempt at a reflecting endpoint. Divided by the weight of the lifted state $(x,i,\epsilon)$, the probability flowing out of it by moves is $r_i^\epsilon$, whereas the flux arriving at it by moves from neighboring configurations equals $r_i^{-\epsilon}$ by Eq.~\eqref{eq:link}. The handoffs and reversals must carry the difference
\begin{equation}
 b_i(x)=r_i^+(x)-r_i^-(x)=\sum_j a_{ij}(x)+c_i(x). \label{eq:split}
\end{equation}
Here the bond currents $a_{ij}=-a_{ji}$ are antisymmetric functions on the bonds that we prescribe, and the residual $c_i$ is the part of $b_i$ that they leave over. The bond current describes how the interaction between $i$ and $j$ should pass activity; the residual is fixed once the currents and the move rates are chosen.

With $[u]_+=\max(0,u)$, a handoff from $i$ to $j$ occurs at rate $[-\epsilon a_{ij}]_+$ and carries the direction $\epsilon$ to $j$. A \emph{reversal} changes $(x,i,\epsilon)$ to $(x,i,-\epsilon)$ at rate $\lambda_{\rm rev}(x,i,\epsilon)=[-\epsilon c_i(x)]_+$. For $c_i>0$, for example, only the negative direction can reverse at $i$. At fixed $x$ and $\epsilon$ the net probability current of handoffs from $i$ to $j$ is $-\epsilon a_{ij}\pi(x)/(2N)$. Adding the contributions of moves, handoffs, and reversals, the incoming minus outgoing probability flux at a lifted state, divided by its weight, is $-\epsilon b_i+\epsilon\sum_j a_{ij}+\epsilon c_i=0$, so the lifted distribution is stationary. Appendix~\ref{app:balance} derives the three terms and also proves global balance for cumulative matching and for a reversible control.

In summary, the sampler repeats the following event.
\begin{enumerate}\setlength{\itemsep}{1pt}
\item In state $(x,i,\epsilon)$, compute the move rate $r_i^\epsilon(x)$, the handoff rates $[-\epsilon a_{ij}(x)]_+$ to the neighbors $j$ of $i$, and the reversal rate $[-\epsilon c_i(x)]_+$.
\item Advance the continuous time by an exponential holding time with the total rate, and select one event in proportion to its rate.
\item Move: change $x_i$ by one step in direction $\epsilon$; an outward attempt at a reflecting endpoint instead flips $\epsilon$. Handoff: set $i\leftarrow j$. Reversal: set $\epsilon\leftarrow-\epsilon$.
\end{enumerate}
The Ising and Potts samplers also refresh the active site at a small rate to keep the chain irreducible (Sec.~\ref{sec:spins}). Section~\ref{sec:pair} adds cumulative matching, which redirects part of the reversals to distant sites.
Appendix~\ref{app:implementation} describes how the implementation tabulates these rates and draws the events efficiently.

\subsection{Move rates from the conditional distribution}\label{sec:conditional}
Each reversal ends a run of same-sign moves, so we choose the move rates that make reversals as rare as possible for prescribed bond currents. Hold all variables except $x_i$ fixed, let $p(t)>0$ be the conditional probability of label $t$, and write $r^\pm(t)$, $b(t)$, and $c(t)$ for the quantities of Sec.~\ref{sec:models} at $x_i=t$. The bond currents enter through $\psi(t)$, the value of $\sum_j a_{ij}$ at $x_i=t$. For the odd bond functions $a_{ij}=A(x_j-x_i)$ used here, $\psi(t)=\sum_{j\sim i}A(x_j-t)$. We assume that its conditional mean $\bar\psi=\sum_t p(t)\psi(t)$ converges absolutely.

Because a move follows the label order in sector $+$ and retraces it in sector $-$, the move rates at site $i$ are fixed by one number per link of the label order. Let $G_t\ge0$ be the conditional probability flux of the forward step $t\to t+1$ in sector $+$. By Eq.~\eqref{eq:link} the backward step $t+1\to t$ in sector $-$ carries the same flux, so the two rates are $G_t/p(t)$ and $G_t/p(t+1)$. Collecting the moves out of $t$ in each sector, $p(t)r^+(t)=G_t$ and $p(t)r^-(t)=G_{t-1}$, so that
\begin{equation}
 p(t)\,b(t)=G_t-G_{t-1}. \label{eq:bflux}
\end{equation}
The ends of the label set fix the boundary values. On a cycle, $G_t$ is periodic. On the integer line, $G_t\to0$ at both ends. On a finite line $0,\ldots,q-1$, $G_{-1}=G_{q-1}=\gamma\ge0$ is the flux of the outward attempts at the two endpoints.

These boundary conditions lead to a lower bound on the reversal rate. With each of them, the sum of Eq.~\eqref{eq:bflux} over $t$ vanishes, so $\av{b}_p=0$. Averaging Eq.~\eqref{eq:split} over $p$ then fixes the conditional mean of the residual, $\av{c}_p=-\bar\psi$, whatever the move rates. The two directions are equally probable at equilibrium, and $[c]_+ + [-c]_+=|c|$, so the mean rate of residual reversals satisfies
\begin{equation}
 \av{\lambda_{\rm rev}}_{p,\epsilon}=\tfrac12\av{|c|}_p\ \geq\ \tfrac12|\bar\psi|. \label{eq:min}
\end{equation}
Equality holds when $c(t)$ has the same sign for all $t$, in particular for the constant residual $c(t)=-\bar\psi$. The moves then absorb the centered part $\psi-\bar\psi$ of the bond currents and leave only the conditional mean $\bar\psi$ to reversals. By Eqs.~\eqref{eq:split} and \eqref{eq:bflux}, this choice requires
\begin{equation}
 G_t-G_{t-1}=p(t)\big[\psi(t)-\bar\psi\big]. \label{eq:G}
\end{equation}

The rates of this construction require a solution of Eq.~\eqref{eq:G} with $G_t\ge0$ on every link; we now show that such a solution exists for each label geometry used here. On the integer line, Eq.~\eqref{eq:G} gives the partial sums $G_t=\sum_{u\le t}p(u)[\psi(u)-\bar\psi]$. For a nonincreasing $\psi$ the centered terms change sign at most once, from positive to negative, and their total vanishes, so every partial sum is nonnegative; $r^+(t)=G_t/p(t)$ is then the discrete inverse Stein operator applied to $\psi$ \cite{steinoperator,stein}. On a cycle, Eq.~\eqref{eq:G} fixes $G_t$ only up to a common constant, the circulating flux, which can be chosen to make every $G_t$ nonnegative for any $\psi$. On a finite line, $G_t=\gamma+\sum_{u=0}^{t}p(u)[\psi(u)-\bar\psi]$, which returns to $\gamma$ at $t=q-1$ and satisfies $G_t\ge\gamma$ for a nonincreasing $\psi$. The outward attempts, at rates $\gamma/p(0)$ in sector $-$ and $\gamma/p(q-1)$ in sector $+$, flip $\epsilon$ at the same site and label. Averaged over $p$ and the directions they add $\gamma$ to the reversal rate, so the minimum total reversal rate at fixed $\gamma$ is $\gamma+|\bar\psi|/2$. The bond currents of the DG, Ising, and Potts models give a nonincreasing $\psi$, and the clock labels form a cycle (Secs.~\ref{sec:dg} and \ref{sec:spins}).

\subsection{A reversal-free chain of integer heights}\label{sec:dg}
The DG model shows when the local construction alone sustains a persistent direction. A periodic hypercubic lattice of spatial dimension $d$ has $N=L^d$ sites and coordination number $D=2d$. Each variable is an integer height $h_i$, and the coupling $K$ penalizes height differences,
\begin{equation}
 \pi(h)\propto\exp\Big[-K\sum_{\langle ij\rangle}(h_i-h_j)^2\Big],
\end{equation}
where the sum counts each bond once. This integer interface model also arises in the description of roughening and its relation to two-dimensional Coulomb gases and periodic Gaussian spin interactions \cite{chui,villain}. Adding the same integer to every height leaves all differences unchanged, so the target distribution is defined on relative heights.

We take $A(u)=2Ku$, so that in the positive sector the activity passes from a higher to a lower site at rate $2K[h_i-h_j]_+$ [Fig.~\ref{fig:transfer}(a)]. Raising a site above its lower neighbors increases the energy of those bonds, and the handoff lets the lower neighbor rise next. To evaluate the residual, write $g_i=Dh_i-\sum_{j\sim i}h_j$. For a candidate height $t$ of site $i$, let $U=Dt-\sum_{j\sim i}h_j$. As $t$ runs over the integers, $U$ runs over $g_i+D\mathbb{Z}$, and the conditional weight is proportional to $\exp(-KU^2/D)$. Since $\psi(t)=-2KU$, the constant residual $c=-\bar\psi$ of Sec.~\ref{sec:conditional} is
\begin{equation}
 c_i=2K\av{U}_p=2K\,\frac{\sum_{U\in g_i+D\mathbb{Z}}U\,e^{-KU^2/D}}{\sum_{U\in g_i+D\mathbb{Z}}e^{-KU^2/D}},
 \label{eq:charge}
\end{equation}
where $\av{U}_p$ is the conditional mean of $U$. The residual therefore depends only on $g_i$ modulo $D$. For $D=2$ each residue class is invariant under $U\mapsto-U$, so $c_i=0$ in every configuration. On the periodic chain, for every $K>0$, the bond handoffs thus balance every finite move in either direction sector, the direction never reverses, and each sector alone preserves the Gibbs distribution of the relative heights [Fig.~\ref{fig:transfer}(b)]. For $D=4$ the even classes have zero residual, and the two odd classes have residuals of equal magnitude and opposite sign. Poisson summation at fixed $D$ gives
\begin{equation}
 c_i=4\pi\, e^{-\pi^2/(DK)}\sin(2\pi g_i/D)
       +O\!\left(e^{-2\pi^2/(DK)}\right), \label{eq:weak}
\end{equation}
so a nonzero residual can be very small at weak coupling. Appendix~\ref{app:obstruction} shows that for $D\geq3$ no nontrivial odd bond function of the height difference, $A(u)=2Ku$ included, removes the residual in \emph{all} environments.

\subsection{Cyclic angles and reflecting spin labels}\label{sec:spins}
The geometry of a variable's state space determines what happens at the ends of a sequence of moves. Clock angles admit a circulating flux around the cycle. For Ising and Potts spins we order the color labels on a line and reflect at its ends.

In the six-state clock model, a unit move rotates an arrow by $\epsilon\pi/3$. Write $\theta_i=2\pi s_i/6$ for $s_i=0,\ldots,5$. The target is
\begin{equation}
 \pi(s)\propto\exp\Big[K\sum_{\langle ij\rangle}\cos(\theta_i-\theta_j)\Big].
\end{equation}
We choose $A(u)=K\sin(2\pi u/6)$ with the difference $u$ measured in clock steps. The conditional law is $p(t)\propto\exp[K\sum_{j\sim i}\cos(2\pi(t-s_j)/6)]$. For any cyclic solution $\widetilde G$ of Eq.~\eqref{eq:G}, we take $G_t=\widetilde G_t-\min_u\widetilde G_u+\eta\min_u p(u)$ with $\eta=0.05$. Subtracting the minimum makes every link flux nonnegative. The positive floor then adds $\eta\min_u p(u)/p(t)$ to both rates at $t$. On a cycle this extra flux circulates without reversals, and it preserves the residual. It is needed for irreducibility: when the local field and all sine currents vanish, the cyclic solution is constant, so without the floor no move, handoff, or reversal would have a positive rate, and such configurations exist on periodic lattices.

For Ising and three-state Potts spins, we use the Potts convention $\pi(s)\propto\exp(K\sum_{\langle ij\rangle}\delta_{s_i,s_j})$ with $q=2$ or 3, so that $K=2\beta J$ for Ising spins with exchange $J$ at inverse temperature $\beta$. We order the labels $0,\ldots,q-1$ on a reflecting line and set $A(u)=Ku/2$ for their integer difference. Here $\psi(t)$ decreases with $t$, so its centered partial sums are nonnegative for every neighbor configuration. Global balance holds for any endpoint flux $\gamma\ge0$ of Sec.~\ref{sec:conditional}, and exact enumeration of small lattices gives the same communicating classes for $\gamma=0$ as for the value used here. We take $\gamma=\eta\min_t p(t)$ with $\eta=0.05$, which keeps every move attempt rate positive. Scaling with $\min_t p(t)$ bounds the endpoint reversal rates $\gamma/p(0)$ and $\gamma/p(q-1)$ by $\eta$ in every local environment, even when an endpoint label is improbable. Unlike the circulating floor of the clock model, these endpoint reflections are reversals. In short test runs on small lattices, $\eta=0$ reduced the number of reversals further and shortened the integrated autocorrelation time of the structure factor $S_1$ (Sec.~\ref{sec:statistics}) of the Ising chains by up to about 40\%, while the Potts chains were unaffected; all benchmarks below use $\eta=0.05$. Handoffs occur only across bonds with unequal labels, and without a further event the lifted chain can be reducible: exact enumeration of the four-site Ising ring finds a closed class that contains the antiferromagnetic configuration. The Ising and Potts samplers therefore also move the activity, at rate 0.01 and with $\epsilon$ kept, to a uniformly drawn site. This refresh preserves the lifted distribution, in which $i$ is uniform (Appendix~\ref{app:balance}), and with it the same enumeration on small periodic lattices with two to four labels in one to three dimensions reaches every configuration and active site. The DG and clock samplers use no refresh; their move rates are positive at every label, in the clock model through the floor $\eta$. In 1D the couplings set the infinite-chain correlation length to eight, and in 2D they are the critical couplings $K=\log(1+\sqrt q)$. In 3D we use the Ising critical point and the three-state Potts coexistence point (Fig.~\ref{fig:gain}).

\section{Cumulative matching of residual currents}\label{sec:pair}
The nearest-neighbor rule is a complete sampler. On a lattice with more than two neighbors per site, however, the minimum in Eq.~\eqref{eq:min} generally leaves nonzero residuals, and each reversal interrupts a chain of same-sign moves. Cumulative matching adds a transfer channel that replaces part of these reversals while keeping the move rates and the bond currents fixed.

We call an event selected by the reversal channel, at rate $[-\epsilon c_i]_+$, a \emph{reversal candidate}. The idea is to pair residuals of opposite sign. At a fixed configuration $x$, a site with $c_i>0$ reverses in the negative sector, and a site with $c_j<0$ reverses in the positive sector. Matching equal portions of the two residuals replaces these two opposite reversal currents by a pair of transfers, one in each sector, so that the activity jumps between the two sites instead of reversing. The construction needs only the list of signed residuals, a site order, and a search window $W$, so the same search applies to all four model families.

Fix a cyclic site order. In one dimension we follow the chain. In higher dimensions we use a Hilbert order at power-of-two lengths \cite{hilbert} and a serpentine order otherwise. Along the periodically continued order, let $H$ be the running sum of the residuals $c_i$, and call a value of $H$ a level. Each site with $c_i>0$ spans an upward interval of levels of $H$, and each site with $c_i<0$ a downward interval [Fig.~\ref{fig:transfer}(c)].

At a reversal candidate, draw a level uniformly in the interval of the source site. From a positive site $i$, search forward to the first site at which $H$ descends through this level; from a negative site, search backward to the first rise. Search at most $W$ other sites. If a recipient is found, transfer the activity to it with $\epsilon$ unchanged; otherwise, reverse. The search leaves the physical configuration unchanged. For each level matched between $i$ and $j$, the intervening path of $H$ lies on one side of that level, so the reverse search from $j$ recovers $i$ within the same window. Let $\ell_{ij}=\ell_{ji}$ be the measure of the levels for which a search from $i$ ends at $j$, and hence a search from $j$ ends at $i$. Define
\begin{equation}
 d_{ij}=\operatorname{sgn}(c_i)\ell_{ij}=-d_{ji},
 \qquad c'_i=c_i-\sum_j d_{ij}. \label{eq:pair}
\end{equation}
The transfer rate from $i$ to $j$ is $[-\epsilon d_{ij}]_+$, and the remaining reversal rate is $[-\epsilon c'_i]_+$. A single source can supply several recipients. Because $d_{ij}$ is antisymmetric, it enters the balance of Eq.~\eqref{eq:split} exactly like an additional bond current between the possibly distant sites $i$ and $j$, and Appendix~\ref{app:balance} verifies global balance for every fixed window. The Ising and Potts samplers retain their endpoint reflections and the independent refresh of the active site.

Matching cannot remove the net residual. At fixed $x$, averaging over the uniform active site and direction gives the mean reversal rate with matching, $(2N)^{-1}\sum_i|c'_i(x)|$. Antisymmetry preserves the signed sum $C(x)=\sum_i c_i(x)$, so
\begin{equation}
 \sum_i |c'_i(x)|\ \geq\ |C(x)|. \label{eq:net}
\end{equation}
With the full window $W=N-1$, the search covers the whole cycle, every residual of the sign opposite to $C$ finds a partner, and ideal matching attains the bound. The implementation represents the transferable residual magnitudes by integer levels and leaves any rounding remainder in the reversal rate. A shorter search can leave opposite residuals unmatched. The bound concerns reversal rates at a fixed configuration; how a handoff affects a long-wavelength fluctuation also depends on where the activity lands.

Each reversal candidate costs a search over at most $W$ sites. In the spin models, including clock, we use $W=\min(64,N-1)$. For square-lattice DG at $K=0.6645$ we use $W=\min(256,N-1)$, and for the other multidimensional DG runs we use the full window $N-1$. Appendix~\ref{app:implementation} describes the sampling cost and the event implementation.

\section{Computational methods}\label{sec:methods}
\subsection{Reference samplers}
We compare the event chains with four single-site samplers, each of which attempts updates at uniformly random sites at total rate one, including rejections. Metropolis proposes an increment of $\pm1$ with equal probability for a height or an angle, and a uniformly chosen different color for an Ising or Potts spin. Heat bath draws from the single-site conditional law. It enumerates the clock or spin states, and uses offsets from $-16$ to $16$ about the floor of the DG conditional center; the discarded probability is below double-precision roundoff at these couplings.

The Suwa--Todo algorithm places the conditional weights consecutively around a circle, in descending order with ties broken by state label, and shifts the source intervals by the largest weight \cite{suwa}. The destination is drawn from the resulting overlap lengths. For two Ising states its transition probabilities coincide with those of Metropolis, although the cost per attempt can differ. IMGS uses the Metropolized-Gibbs proposal with maximal skewness ($\delta=1$), with the change of the selected height or integer label as its guide \cite{faizi}. If $a_\epsilon$ is the probability of a move averaged over all site choices at fixed configuration in sector $\epsilon$, a rejected attempt switches direction with probability $[a_{-\epsilon}-a_\epsilon]_+/(1-a_\epsilon)$.

\subsection{Effective sample size, computational cost, and relaxation}\label{sec:statistics}
For each model we measure the bond energy and the longest-wavelength structure factor,
\begin{equation}
 S_1=\frac{1}{dN}\sum_{\mu=1}^d\left|\sum_j \phi_j
      \exp(2\pi\mathrm{i} x_{j,\mu}/L)\right|^2 ,
\end{equation}
where $x_{j,\mu}$ is the coordinate of site $j$ along axis $\mu$. For DG, $\phi_j=h_j$. For clock spins, $\phi_j=(\cos\theta_j,\sin\theta_j)$ and the squared norm sums the Fourier powers of the two components. For Ising and Potts spins, we use the complex scalar $\phi_j=\exp(2\pi\mathrm{i}s_j/q)$. The resulting $S_1$ is insensitive to a uniform height shift or a common spin rotation. We also record the DG height variance $W_h^2=N^{-1}\sum_j(h_j-\bar h)^2$ and its square. For spins, $|m|^2$ is the squared norm of the spatial mean of $\phi_j$.

We measure time in \emph{event sweeps}. One event sweep comprises $N$ elementary events: moves, handoffs, reversals, or refreshes of the active site for an event chain, and attempted updates for a random-site sampler. A reversal thus costs an event even though the configuration does not change. We record observations at independent Poisson times. If $n$ observations span $n_e$ events, and $\tau_s=\tfrac12+\sum_{k\ge1}r_k$ is the integrated time of their normalized lag correlation $r_k$, the autocorrelation time in event sweeps and the effective sample size (ESS) of the recorded mean are
\begin{equation}
 \tint=(\tau_s-\tfrac12)\frac{n_e}{Nn},\qquad
 {\rm ESS}=\frac{n}{2\tau_s}. \label{eq:tau}
\end{equation}
Because observations are taken at Poisson times independent of the dynamics, $\tau_s-\tfrac12$ estimates the continuous-time correlation integral in units of the mean observation spacing, which $n_e/(Nn)$ converts to event sweeps, whereas the variance of the sample mean involves $\tau_s$ itself.

The production rate is the ESS summed over replicas and divided by their combined production core-seconds, including measurements. For each model and size, Fig.~\ref{fig:gain} compares the six methods using the same executable and recording interval. Each trajectory used one CPU core. The C++17 programs were compiled with \texttt{g++ -O3 -march=native}. Initialization and burn-in are excluded from production time. Independent measurements at other recording intervals can resolve $\tint$ at additional sizes.

We call a size \emph{admitted} when all diagnostics of Appendix~\ref{app:statistics} pass. Confidence intervals are given per comparison (individual), without multiple-comparison correction. Comparisons between methods use a common mean recording interval $\Delta$. When $\Delta$ is comparable to or longer than the relaxation times, nearly every observation is already independent, and the rate ratio mainly reflects the cost per event; it then understates the advantage of the faster-relaxing method. We therefore also compare $\tint$ in event sweeps, each estimated at the finest admitted recording interval of its method. Ratios of these times count elementary events and do not depend on the implementation or the machine, whereas production rates include the cost of each event and of the measurements. Each reported time requires at least four independent replicas and a total $\mathrm{ESS}\ge1000$, together with correlation-window and equilibration checks (Appendix~\ref{app:statistics}). At Potts coexistence we also require visits to both phases. We fit $\log\tint=\log A+z\log L$ separately for each observable and method, and for each site order and window regime of cumulative matching.

\section{Results}
\subsection{Relaxation times and production rates}\label{sec:throughput}
The slowest observable of the discrete interface is its longest-wavelength structure factor $S_1$. Near roughening ($K=0.6645$) on a $64\times64$ lattice, the four random-site samplers produce 10.2--12.8 effective samples of $S_1$ per core-second. Nearest-neighbor lifting produces 295, about 23 times the fastest of them. Cumulative matching raises the rate to 733, a further factor of $2.49$ (individual approximate 95\% interval $2.39$--$2.58$). These rates include the time spent finding the next active site and measuring the observable. Measured in elementary events, $\tint(S_1)$ is 186--386 event sweeps for the random-site samplers, 19.2 for nearest-neighbor lifting, and 4.36 with cumulative matching.

\begin{figure*}[!t]
\centering\includegraphics[width=\linewidth]{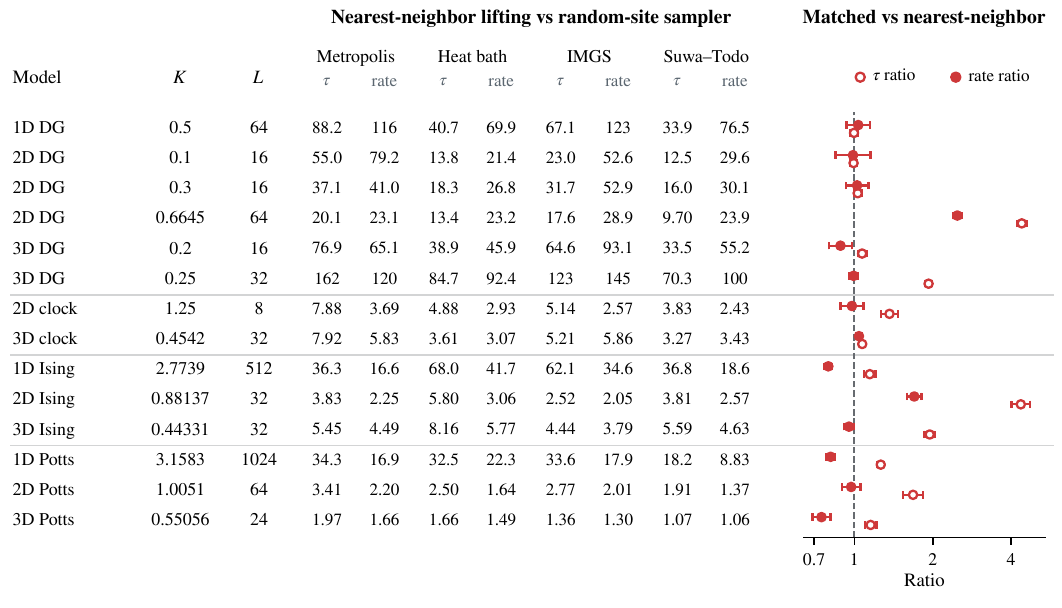}
\caption{Relaxation-time and production-rate ratios for $S_1$. Left: for each named random-site sampler, the column $\tau$ divides its $\tint(S_1)$ by that of nearest-neighbor lifting, both in event sweeps, and the column ``rate'' divides the rate of nearest-neighbor lifting in effective samples per core-second by that of the sampler. Right: the same two ratios for cumulative matching relative to nearest-neighbor lifting, as open ($\tau$) and filled (rate) circles with individual approximate 95\% intervals. A ratio above one favors nearest-neighbor lifting on the left and matching on the right. Each row uses the largest $L$ at which all 15 method pairs pass the diagnostics at a common recording interval for $S_1$; each time is estimated at the finest admitted interval of its method. The 1D DG runs with and without matching have identical dynamics and independent seeds. Couplings $K$ follow the conventions in Sec.~\ref{sec:models}. A recording interval comparable to or longer than the relaxation times understates the rate advantage of the faster-relaxing method (Sec.~\ref{sec:statistics}).}\label{fig:gain}
\end{figure*}

Figure~\ref{fig:gain} extends both comparisons to every model, each at the largest size that admits all six methods at a common recording interval. For each random-site sampler the table gives the ratio of its $\tint(S_1)$ to that of nearest-neighbor lifting and the ratio of the production rates, and the right panel gives the same two ratios for cumulative matching relative to nearest-neighbor lifting. In every row nearest-neighbor lifting has a shorter relaxation time and a higher measured rate than every random-site sampler, although at Potts coexistence on the cubic lattice both ratios relative to Suwa--Todo, 1.07 and 1.06, have intervals that include one. The two ratios differ systematically. For DG the rate ratio exceeds the time ratio in all entries except those relative to Metropolis in 3D, so in this implementation an elementary event of nearest-neighbor lifting usually costs less than an attempted random-site update. For the clock, Ising, and Potts models the time ratios are larger; part of this difference also reflects recording intervals comparable to the relaxation times. Matching never lengthens $\tint(S_1)$ beyond its uncertainty and shortens it by up to a factor of 4.4. Its gain in rate is smaller wherever it shortens the time appreciably, because each reversal candidate costs a search. Relative to the four random-site samplers, the event chain with cumulative matching shortens $\tint(S_1)$ by factors of 43--89 near roughening, 5.2--11 in the square-lattice clock model, and 11--25 and 3.2--5.7 at the square-lattice Ising and Potts critical points, and it raises the production rate by factors of 58--72, 2.4--3.6, 3.5--5.2, and 1.3--2.1, respectively.

At the square-lattice Ising critical point ($L=32$), matching shortens $\tint(S_1)$ by a factor of 4.4 and adds a factor of $1.70$ to the rate of nearest-neighbor lifting. At the three-state Potts critical point ($L=64$) it shortens the time by a factor of 1.7, but the two event chains have similar rates. Whether matching pays off depends on the model and the size, and Fig.~\ref{fig:ratebysize} follows the two event chains to additional sizes at which their rates can be compared directly. The following subsections examine the models in turn.

\subsection{Persistent transport in the discrete Gaussian chain}

\begin{figure*}[!t]
\centering\includegraphics[width=\linewidth]{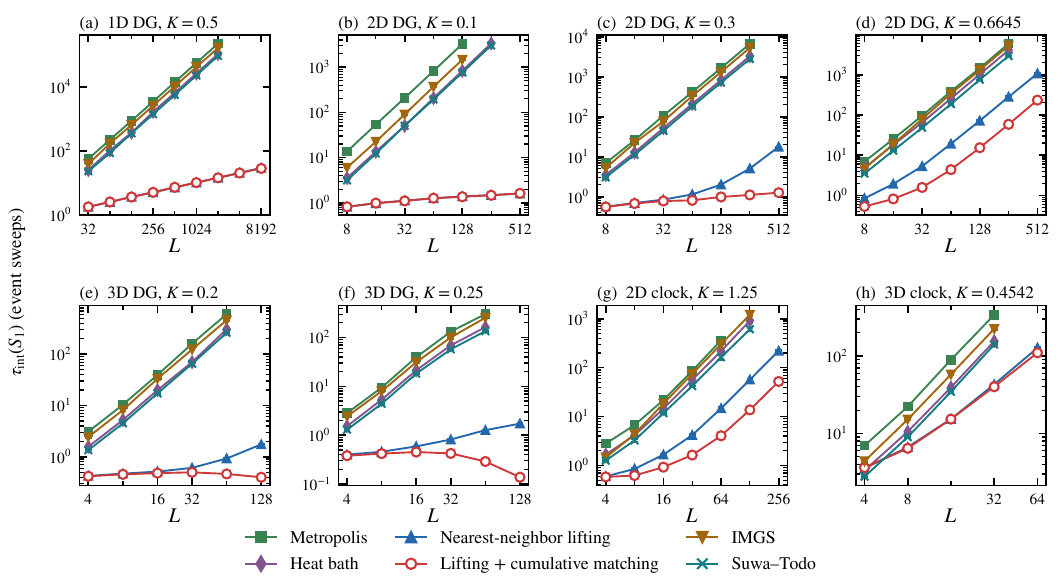}
\caption{Integrated autocorrelation time of the longest-wavelength structure factor $S_1$ in event sweeps ($L^d$ elementary events per sweep). Error bars show one standard error. Lines connect successive measured sizes for each method. In (a) nearest-neighbor lifting and lifting with cumulative matching have identical dynamics and independent seeds, so the blue symbols are hidden behind the red ones. In (d) the matching window is $W=N-1$ for $L\le16$ and $W=256$ for larger $L$. Table~\ref{tab:structure} gives the fitted slopes and size ranges.}\label{fig:sizes}
\end{figure*}

The one-dimensional DG model isolates the effect of the nearest-neighbor chain, because its residual vanishes identically [Eq.~\eqref{eq:charge}] and matching has nothing to act on. For $S_1$ at $K=0.5$ the fitted size exponent of $\tint$ is $0.507\pm0.016$ over $L=32$--256 [Fig.~\ref{fig:sizes}(a)]. Independent runs with matching enabled sample the same dynamics and give $0.503\pm0.016$. The four random-site samplers have slopes near two, including $2.003\pm0.030$ for IMGS and $1.986\pm0.015$ for Suwa--Todo (Table~\ref{tab:structure}). Here and below, slope uncertainties are approximate 95\% half-widths over the stated fit range. The adjacent-size slopes of nearest-neighbor lifting stay between 0.47 and 0.53 up to $L=8192$.

At $L=64$ nearest-neighbor lifting produces effective samples of $S_1$ about 70 times as fast as heat bath, the fastest random-site sampler at that size. The mechanism is the one sketched in Fig.~\ref{fig:transfer}(b): the active site moves along the chain while raising heights in the positive sector or lowering them in the negative sector, so a Fourier amplitude at wavelength $L$ can change coherently over many successive moves. A trajectory measurement at $L=1024$ is consistent with this picture. The mean-square change of $S_1$ grows nearly quadratically with the separation in events over short and intermediate lags, as for ballistic motion. Symmetrizing the generator with its stationary time reversal, at unchanged escape rates, removes the stationary current and gives nearly linear, diffusive growth over the same range [Fig.~\ref{fig:transport}(d)]. This comparison probes the role of the stationary current within the nearest-neighbor sampler, while the exponent fit describes the observed relaxation of the height mode.

\subsection{Residual currents on square and cubic lattices}
Square-lattice DG lets us vary the strength of the residual within the same bond-current construction. At $K=0.1$ and 0.3 the interface is rough. The coupling $K=0.6645$ is near the roughening transition, for which estimates include $0.6645(6)$ and $0.6653(2)$ \cite{rough,rough2}. At $K=0.1$ the residual amplitude is only about $2.4\times10^{-10}$ [Eq.~\eqref{eq:weak}]. Nearest-neighbor lifting already relaxes rapidly, and the two lifted curves nearly coincide [Fig.~\ref{fig:sizes}(b)].

In the rough phase the event chains are expected to show a crossover. At small sizes a chain of same-sign moves can sweep the long-wavelength mode before a reversal occurs, and $\tint(S_1)$ grows only weakly with $L$. Once residual reversals randomize the direction on the scale of that mode, the growth should cross over to the diffusive $L^2$ behavior of the random-site samplers. The crossover size should increase as the residual weakens, and matching, which removes part of the reversals, should push it to larger $L$. At $K=0.1$ the two lifted chains have slopes of $0.13\pm0.02$ and $0.13\pm0.03$ up to $L=256$ (Table~\ref{tab:structure}), so the crossover lies beyond the simulated sizes.

\begin{figure*}[!t]
\centering\includegraphics[width=\linewidth]{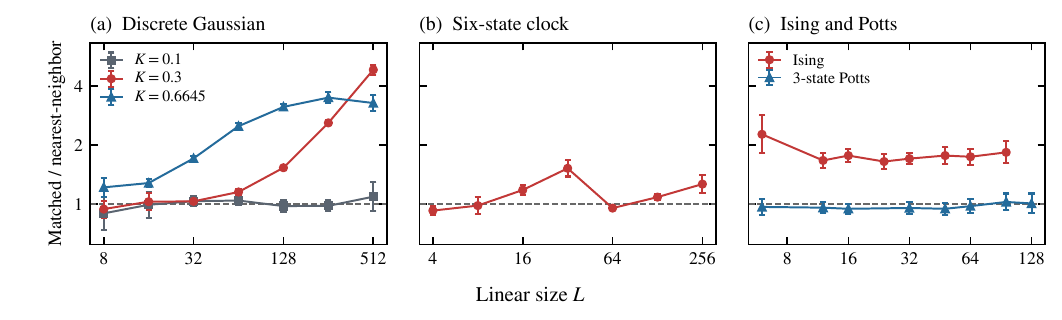}
\caption{Effect of cumulative matching on effective samples per production core-second in the square-lattice models. Each point divides the rate with matching by that of nearest-neighbor lifting at a common recording interval, so a ratio above one favors matching. The observable is $S_1$ in all panels. Lines connect successive admitted sizes at each coupling. Bars give individual approximate 95\% intervals.}\label{fig:ratebysize}
\end{figure*}

At $K=0.3$ the crossover is visible for nearest-neighbor lifting, and matching moves it beyond the simulated sizes [Fig.~\ref{fig:sizes}(c)]. The adjacent-size slopes of $\tint(S_1)$ for nearest-neighbor lifting rise from 0.81 ($L=64$--128) to 1.33 (128--256) and 1.81 (256--512), approaching the random-site value of two. With matching the slope is $0.176\pm0.016$ on $L=16$--256, and this fit predicts the admitted point at $L=512$. At that size matching lowers $\tint(S_1)$ from 17.78 to 1.264 event sweeps, a factor of 14.1. Counting search and measurement time, the gain in production rate is $4.83$ ($4.54$--$5.13$), compared with $1.15$ at $L=64$ and $2.59$ at $L=256$ [Fig.~\ref{fig:ratebysize}(a)]. The comparison that requires all six methods at a common interval reaches only $L=16$ at this coupling.

Near roughening, matching reduces reversals, relaxation times, and production cost by different factors. At $L=64$, direct trajectory counts give $38\,478$ reversals for nearest-neighbor lifting and $4\,043$ with matching in $1\,048\,576$ events per method. In independent equilibrium runs $\tint(S_1)$ falls from 19.17 to 4.36 event sweeps, consistent in direction with the gain in production rate in Sec.~\ref{sec:throughput}. At $L=256$, the times are 280.7 and 58.3 event sweeps and the rate gain is $3.48$ ($3.28$--$3.70$). Both curves steepen at large sizes, and the large-size fit with matching retains curvature (Table~\ref{tab:structure}).

The cubic-lattice DG model has a finite equilibrium correlation length at fixed $K>0$ \cite{gopfert}. At $K=0.2$ and $0.25$ nearest-neighbor lifting already outperforms all four random-site methods for $S_1$ at the common sizes (Fig.~\ref{fig:gain}). Matching costs throughput at $K=0.2$, $L=16$, but at $L=128$ its rate relative to nearest-neighbor lifting is $1.14$ ($1.04$--$1.25$) and $1.56$ ($1.44$--$1.70$) at the two couplings. As in the rough phase of the square-lattice model, the size dependence is best read as a crossover [Fig.~\ref{fig:sizes}(e,f)]. For nearest-neighbor lifting at $K=0.2$, the adjacent-size slopes of $\tint(S_1)$ are 0.13--0.24 up to $L=32$ and rise to 0.61 ($L=32$--64) and 0.91 (64--128), while the random-site samplers have slopes of 1.90--2.13 between $L=32$ and 64. With matching, $\tint(S_1)$ stays below 0.5 event sweeps up to $L=128$ at both couplings, whereas the random-site times grow by factors of 100--200 between $L=4$ and 64. At $K=0.25$ the random-site slopes fall to 1.2--1.3 between $L=32$ and 64, consistent with the approach to the finite correlation length. The negative fitted slopes with matching (Table~\ref{tab:structure}) describe only this finite range.

\subsection{Ising and Potts spins with reflecting labels}

In one dimension the event chain helps even when the slow mode has a size-independent relaxation time. The Ising and Potts couplings give a correlation length $\xi=8$ through $\xi^{-1}=\log[(e^K+q-1)/(e^K-1)]$. Once $L\gg\xi$, the random-site times for $S_1$ in sweeps approach a constant. Nearest-neighbor lifting nevertheless samples $S_1$ 16.6--41.7 (Ising, $L=512$) and 8.8--22.3 (Potts, $L=1024$) times as fast as the four random-site methods. Adding the matching search lowers the measured rates there by factors of 0.79 and 0.81.

\begin{figure*}[!t]
\centering\includegraphics[width=\linewidth]{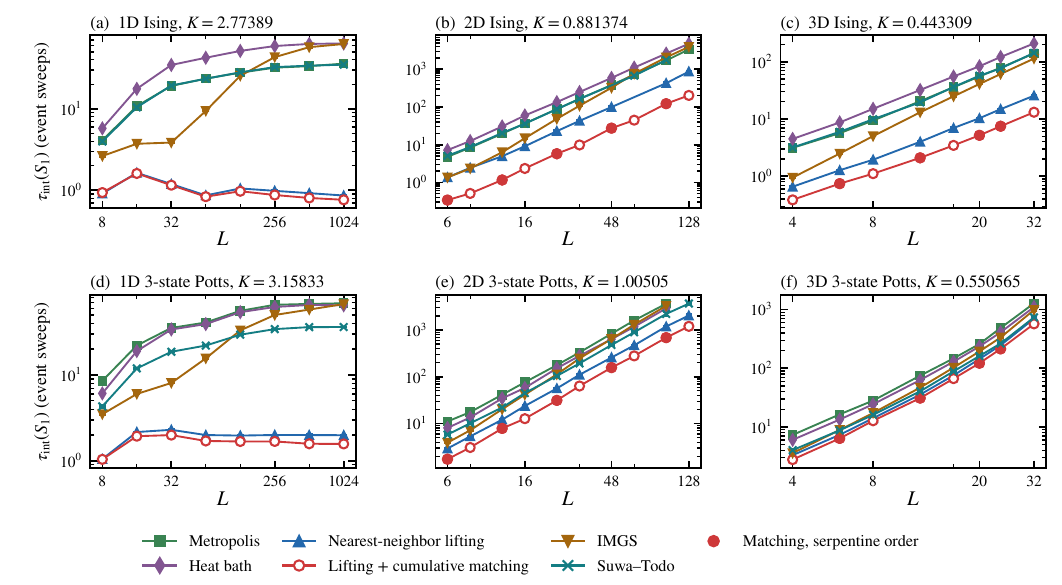}
\caption{Integrated autocorrelation time of the longest-wavelength structure factor $S_1$ for Ising (top row) and three-state Potts (bottom row) models in 1D, 2D, and 3D (columns). The red line connects successive measured sizes with cumulative matching, which uses the full window $W=N-1$ for $N\le65$ and $W=64$ otherwise; filled red circles indicate serpentine site order, and open red circles linear or Hilbert order, so in 2D and 3D one line joins both orders. All points satisfy the observable-specific diagnostics. Bars show one standard error.}\label{fig:spins}
\end{figure*}

At the square-lattice critical points the event chains lower the prefactor of $\tint(S_1)$ rather than its size exponent. Apart from the steeper descriptive slopes of IMGS (2.51 and 2.36), the fitted exponents are 2.13--2.21 for Ising and 2.07--2.16 for Potts, although most of these fits end at the data boundary or are descriptive and their ranges differ between methods (Table~\ref{tab:spinmodes}). At the six-method sizes of Fig.~\ref{fig:gain}, nearest-neighbor lifting samples $S_1$ 2.05--3.06 times as fast as the four random-site methods for Ising ($L=32$) and 1.37--2.20 times as fast for Potts ($L=64$). At $L=96$, where matching uses the serpentine order because $L$ is not a power of two, it shortens $\tint(S_1)$ from 428.3 to 122.4 event sweeps for Ising and from 1187 to 686 for Potts. For Ising this raises the production rate by $1.83$ ($1.61$--$2.08$) [Fig.~\ref{fig:ratebysize}(c)], and at $L=32$ the matched rate is 3.48 times that of IMGS and 4.37 times that of Suwa--Todo. At $L=32$ the common recording interval, $\Delta=51.3$, exceeds both $\tint(S_1)$ values, 42.9 and 9.85 event sweeps, so the factor $1.70$ there understates the effect of the fourfold shorter time. For Potts the search cost offsets the shorter correlation time, and the rate ratio stays close to one at all sizes, $1.02$ ($0.92$--$1.13$) at $L=96$. The ratios to nearest-neighbor lifting isolate the effect of matching, because within each model the two event chains share the same move rates, bond handoff rates, and endpoint reflections.

In three dimensions nearest-neighbor lifting remains the fastest of the six methods, but matching does not raise the measured rate. At the Ising critical point, $L=32$, nearest-neighbor lifting samples $S_1$ 3.79--5.77 times as fast as the random-site methods. Matching halves $\tint(S_1)$, from 25.6 to 13.2 event sweeps, but the common recording interval $\Delta=34.3$ is longer than both times, and the rate ratio, $0.953$ ($0.911$--$0.997$), mainly reflects the search cost. At three-state Potts coexistence, $L=24$, nearest-neighbor lifting has the highest measured rate, 1.06--1.66 times those of the random-site methods, and matching changes it by a factor of $0.748$ ($0.691$--$0.809$); at $L=32$ the direct ratio is $0.695$ ($0.612$--$0.789$). Because the chains must visit coexisting phases and the fitted slopes drift, the finite-size Potts slopes near three should not be interpreted as continuous-transition exponents (Table~\ref{tab:spinmodes}).

\subsection{Relaxation of orientational fluctuations}
The square-lattice six-state clock model at $K=1.25$ ($T=0.8$ in units of the exchange) lies between its two Berezinskii--Kosterlitz--Thouless transitions \cite{jose,tomita}. At the common size $L=8$, nearest-neighbor lifting yields about 2.4--3.7 times the $S_1$ rates of the four random-site methods, and matching gives a comparable rate. At $L=256$ the advantage of matching is clear for both $S_1$ (220.3 to 52.2 event sweeps) and $|m|^2$ (170.8 to 44.5). The corresponding production rate ratios, $1.26$ ($1.14$--$1.39$) and $1.14$ ($1.03$--$1.27$), include the search cost. The rate ratio for $S_1$ does not grow monotonically with size [Fig.~\ref{fig:ratebysize}(b)]: the dip at $L=64$ coincides with a jump of the common recording interval from 1.20 to 52.3, longer than both relaxation times there (14.7 and 4.06 event sweeps). The energy relaxes faster than either magnetic observable at the larger sizes (Fig.~\ref{fig:clockplane}).

In the cubic-lattice clock model at $K=0.4542$, near its ordering point \cite{zou}, sixfold anisotropy is irrelevant at the three-dimensional XY critical fixed point but influences the ordered phase \cite{shao}. At $L=32$ the $S_1$ rate of nearest-neighbor lifting is 5.86 times that of IMGS. Matching brings the ratio to 6.10, an additional factor of only $1.041$ ($1.016$--$1.067$). At $L=64$ even the matched chain takes 339.9 event sweeps for the uniform amplitude $|m|^2$, compared with 110.1 for $S_1$ (Fig.~\ref{fig:clock}). The four-point fits of Table~\ref{tab:clockslopes}, 1.87--1.88 for $S_1$ with the random-site samplers over $L=4$--32, are lowered by the smallest sizes. Between adjacent sizes their slopes rise from 1.66--1.79 ($L=4$--8) to 1.93--2.02 (16--32), consistent with a dynamic exponent near two. Over the same range $L=16$--32 the event chains have local slopes of 1.48 (nearest-neighbor lifting) and 1.39 (matching). These still increase, to 1.57 and 1.46 between $L=32$ and 64, so the asymptotic exponent of the event chains is not resolved.

\section{Discussion and conclusions}
The single-site conditional distribution contains the energy change for every candidate value of the active variable. Combined with antisymmetric currents on the bonds, these weights determine both a finite-step update and a rule for choosing the next active variable. Centering the prescribed bond current over the conditional law minimizes the mean reversal rate on a line or a cycle, and on a finite line the endpoint flux adds a known contribution to the total reversal rate. On a one-dimensional DG interface the minimum is zero in every environment, so the same direction persists across arbitrarily many handoffs. The relaxation exponent near one half in this case belongs to the nearest-neighbor algorithm itself.

Cumulative matching addresses the imbalance that the local construction leaves behind. It changes the destination of a residual event but not the move rates. Even a full search preserves the net signed residual of the configuration [Eq.~\eqref{eq:net}], which limits how many reversals matching can remove. Two further effects separate fewer reversals from faster sampling. A unit move at site $i$ changes the longest-wavelength height amplitude by $\epsilon\cos(2\pi x_{i,1}/L)$. After a distant handoff this factor can change sign, so the mode can move backward even though $\epsilon$ is unchanged. In addition, inspecting sites costs time. Square-lattice DG near roughening at $L=64$ illustrates both effects: matching reduces the reversal count by a factor of about 9.5, $\tint(S_1)$ by 4.4, and the cost per effective sample by 2.5. These three factors come from separate trajectory, equilibrium, and production measurements.

In the spin benchmarks matching shortens $\tint(S_1)$ in every model, by factors from 1.15 to 4.4, but whether it pays off in computing time depends on the model. In square-lattice Ising, matching nearly doubles the rate of nearest-neighbor lifting. In square-lattice Potts, matching shortens $\tint(S_1)$ but does not raise the rate per core-second. At Potts coexistence on the cubic lattice it lowers the rate, and at the cubic-lattice Ising critical point the comparison is limited by a recording interval longer than both relaxation times. In every case nearest-neighbor lifting alone has a shorter relaxation time and a higher measured rate than the four random-site samplers at the largest size common to all six methods, although at Potts coexistence on the cubic lattice the intervals of both ratios relative to Suwa--Todo include one. Because only the comparison between the two event chains holds the move and nearest-neighbor handoff rates fixed, it tests directly whether fewer reversals justify the search cost for a given observable.

The benchmarks correspond directly to surface and adsorption problems. In the DG model the integer at each site is the height of a surface column \cite{chui}, so $S_1$ measures long-wavelength fluctuations of the surface profile. The Ising model is equivalent to a lattice gas of adsorbed particles with nearest-neighbor attraction $2K$ at the particle--hole symmetric chemical potential \cite{leeyang}; with occupations $n_j=s_j$, $S_1$ is four times the occupancy structure factor at the lowest wave number. In molecular simulations, coupled protonation states at fixed pH \cite{protonation} form a similar set of interacting binary variables, whose conditional weights would come from the pH-dependent free energies of each titratable site in its environment. Other candidates include quantum spin--phonon updates \cite{spinphonon} and tensor-network Monte Carlo methods that sample combinations of singular values \cite{tnmc}. Each application needs an order of the local states, their conditional weights, and the currents on the selected bonds, and its benefit should be judged by the relaxation of the target observable, including the cost of each handoff.

\section*{Acknowledgments}
Calculations were performed using computational resources of the Supercomputer Center at the Institute for Solid State Physics, the University of Tokyo. This work was supported by the Center of Innovation for Sustainable Quantum AI (SQAI), JST under Grant No. JPMJPF2221 and by CREST under Grant No. JPMJCR24I1. The author acknowledges the use of AI tools from OpenAI and Anthropic for brainstorming ideas, coding, running numerical simulations, making figures, and editing the text. The author has reviewed all content and takes full responsibility for the results.

\appendix
\numberwithin{equation}{section}

\section{Global balance for nearest-neighbor lifting and its extension}\label{app:balance}
Let $s=(x,i,\epsilon)$ and $\mu(s)=\pi(x)/(2N)$. For transition rates $R(s,s')$ the global balance equation at a single lifted state is
\begin{equation}
 \sum_{s'\ne s}\mu(s')R(s',s)=\mu(s)\sum_{s'\ne s}R(s,s').
 \label{eq:gb}
\end{equation}
Sum the skewed detailed balance condition~\eqref{eq:link} over the configurations $x'$ adjacent to $x$ at site $i$. The incoming physical flux in sector $\epsilon$, divided by $\mu(s)$, equals the total move rate in the opposite sector,
\begin{equation}
 \sum_{x'}\frac{\pi(x')}{\pi(x)}k_i^\epsilon(x',x)
       =\sum_{x'} k_i^{-\epsilon}(x,x').
\end{equation}
At a reflecting endpoint an outward attempt arrives in the opposite sector at the same label, whose target weight is again $\mu(s)$. Including these attempts, the normalized incoming and outgoing fluxes from moves and endpoint reflections are $r_i^{-\epsilon}(x)$ and $r_i^\epsilon(x)$, respectively, and their difference is $-\epsilon b_i(x)$.

Handoffs and residual reversals occur at fixed $x$, so every source and destination of these events has the same $\mu$. Denote their rates by $T^\epsilon_{ij}=[-\epsilon a_{ij}]_+$ and $F_i^\epsilon=[-\epsilon c_i]_+$. Antisymmetry $a_{ji}=-a_{ij}$ and the identity $[u]_+-[-u]_+=u$ give
\begin{align}
 T^\epsilon_{ji}-T^\epsilon_{ij}&=\epsilon a_{ij},\\
 F_i^{-\epsilon}-F_i^\epsilon&=\epsilon c_i .
\end{align}
The normalized total incoming minus outgoing flux is thus
\begin{equation}
 -\epsilon b_i+\epsilon\sum_j a_{ij}+\epsilon c_i=0
 \label{eq:gbcancel}
\end{equation}
by Eq.~\eqref{eq:split}. This proves Eq.~\eqref{eq:gb} for every $x,i,\epsilon$. A uniform refresh of the active site also preserves $\mu$, since it draws from the conditional law of $i$ at fixed $x$ and $\epsilon$. Summing $\mu$ over the auxiliary variables recovers $\pi(x)$. For the integer Gaussian model at fixed $N$, the escape rates grow at most linearly in the height range, which a single move can increase by at most one; comparison with a linear birth process excludes infinitely many events in finite time.

For cumulative matching, each matched level traverses the same distance in the forward and backward searches (Sec.~\ref{sec:pair}), so its contribution to $d_{ij}$ has the opposite sign to its contribution to $d_{ji}$. The transfer channels therefore contribute $\epsilon\sum_j d_{ij}$ to Eq.~\eqref{eq:gbcancel}, exactly canceling the change $\epsilon(c'_i-c_i)$ from reversals. The move rates are untouched. The argument applies to each representable level within any fixed search window.

Finally, the stationary time reversal has off-diagonal rates $R^\dagger(s,s')=\mu(s')R(s',s)/\mu(s)$. Its outgoing traffic is the incoming traffic of the generator $Q$ with rates $R$, and conversely. Since $Q$ satisfies Eq.~\eqref{eq:gb}, so does $Q^\dagger$, and by linearity so does every mixture $Q_\rho$ of Appendix~\ref{app:transport}. This is a stationarity statement. In the reversal-free one-dimensional height model each direction sector is closed, and the same calculation within either sector gives the Gibbs distribution of the relative heights.

\section{An obstruction for difference-only bond currents}\label{app:obstruction}
Consider an integer variable with conditional weight $\prod_{k=1}^D w(t-y_k)$ for arbitrary integer neighbor values. Let $w$ be positive, even, and summable with a nowhere-vanishing Fourier transform, and assume that $wA$ is summable and $A$ is odd. Suppose that the flux at infinity vanishes and that $c_i=0$ for every environment. Conditional centering then requires
\begin{equation}
 \sum_t\sum_{k=1}^D g(t-y_k)\prod_{\ell\ne k} w(t-y_\ell)=0,
\end{equation}
where $g(v)=w(v)A(-v)$. Set $y_D=0$ by translation and Fourier transform in the other $D-1$ independent differences. With $\widehat w(\theta)=\sum_v w(v)e^{\mathrm{i}v\theta}$ and the same convention for $g$, the sum over the common translated coordinate imposes $\theta_D=-\sum_{k<D}\theta_k$ modulo $2\pi$. On that set the transform gives
\begin{equation}
 \sum_{k=1}^{D}\widehat g(\theta_k)\prod_{\ell\ne k}\widehat w(\theta_\ell)=0 .
\end{equation}
Division by the nonzero product $\prod_\ell\widehat w(\theta_\ell)$ gives $\sum_k F(\theta_k)=0$ for $F=\widehat g/\widehat w$, a continuous odd function on the circle with $F(0)=0$. For $D\geq3$ take the angles $u,v,-u-v$ with all other angles zero. Then $F(u+v)=F(u)+F(v)$. Since $F$ is bounded on the circle and $F(nu)=nF(u)$, it vanishes everywhere, and thus $A=0$. A vanishing residual would then make the physical link flux constant, hence zero by its boundary condition. The Gaussian bond weight satisfies the transform hypothesis by Poisson summation. For $D=2$ there are only two frequencies, $\theta$ and $-\theta$, so oddness alone satisfies the transformed condition; this is why the integer chain can have nonzero handoff currents and zero residual. On a finite cyclic state space the same argument works when the discrete transform of $w$ has no zeros, but a constant physical link flux can then circulate around the cycle even when the bond current vanishes. The obstruction concerns a single common function of a bond difference. The configuration-dependent transfers of Sec.~\ref{sec:pair} lie outside this class.

\section{Event implementation and random-site controls}\label{app:implementation}

We call a run of consecutive moves at one site, ended by the first other event, a \emph{residence}. If the move, the handoffs, and the reversal have total rate $\Lambda$, an implementation can draw an exponential holding time of mean $1/\Lambda$ and select a channel in proportion to its rate. An independent Poisson observation clock can be included among these channels, so that the observation rate is independent of the current configuration even when $\Lambda$ depends strongly on the neighbors. Recording the configuration at each jump would instead weight fast-exiting states more heavily.

Our C++ implementation tabulates rates by local environment. For DG it draws a residence at the active site by a Walker alias table \cite{walker}. The separation between constructing a distribution and drawing from it also underlies efficient geometric-allocation samplers \cite{allocationproceedings}. To construct the table, let $p_n$ be the conditional probability of one further forward move after $n$ such moves and $p_n(e)$ the probability of an exit of type $e$, including observation. The joint probability of $n$ moves followed by $e$ is $p_n(e)\prod_{k<n}p_k$. The cumulative displacement, event count, and observables are updated together. If a residence reaches the table cutoff, its entire surviving probability is a continuation outcome. A radix key encodes the oriented height differences in a fixed neighbor order, and alias columns for the environments encountered share a contiguous memory pool. Rare differences beyond the table range use the original competing rates. The clock implementation instead tabulates the local rates by the histogram of the six neighbor angles relative to the active spin, and then selects an actual neighbor with its bond rate.

The matching search uses maps between sites and their ranks in the cyclic order, requiring $O(N)$ storage. DG can cache the gradient residue in curve order: a height move affects only the neighboring residues, since it changes the active site's own gradient by a multiple of $D$. If a fraction $f$ of the elementary events are reversal candidates and the mean scan length per candidate is $n_{\rm scan}$, the extra site inspections cost $f n_{\rm scan}$ per event.

IMGS and Suwa--Todo provide two different alternatives to spatial handoffs. One sweep of a random-site sampler comprises $N$ attempted updates.

The Suwa--Todo algorithm reduces the probability of a rejected update. Other discrete lifted samplers favor the growth or shrinkage of a walk or graph \cite{walkgrowth,isingworm}, and directed worms use flow allocations or lifting to reduce backtracking at the head \cite{worm,liftedworm}. Rejection-control experiments on Potts spins found that the relation between rejection and autocorrelation depends on the site-selection order \cite{rejection}. In the comparisons here all four single-site controls use random sites. A residual reversal in an event chain changes its direction and counts as an elementary event even though the configuration stays fixed.

\section{Correlation windows and finite-size fits}\label{app:statistics}
For observations taken at Poisson rate $1/(N\Delta)$, the sum of the densities of all positive arrival times equals $1/(N\Delta)$. Hence $(\tau_s-\tfrac12)\Delta$ estimates the continuous-time correlation integral in units of $N$ microscopic time units. Substituting the measured event spacing gives Eq.~\eqref{eq:tau}.

We estimate lag covariances by Fourier transform \cite{wolff}. For each recorded series we find a lag exceeding ten times the running sum of absolute correlations, including the half term. We take the longest such crossing among the six recorded series, bounded between 32 and $n/16$. For DG the six series are the bond energy, $W_h^2$, $S_1$, the structure factor at twice the lowest wave number, the real Fourier amplitude at the lowest wave number, and $W_h^4$; for spins they are the bond energy, $|m|^2$, $S_1$, $|m|$, the magnetization projected on one axis, and $|m|^4$. This window avoids early zero crossings of oscillatory modes, and signed correlations enter the integral. We check agreement with batch means, stability across block lengths, and the trajectory length relative to the window.

At a common recording interval we pool sets of independent runs if their times agree within three combined standard errors and their means within five. We use the finest admitted interval for each observable and require agreement across admitted intervals within three combined standard errors. Every time estimate uses at least four independent replicas and a total $\mathrm{ESS}\ge1000$. For Ising and Potts we compare ordered and random starts. At Potts coexistence we require visits to both phases, with four or more crossings between $|m|\le0.10$ and $|m|\ge0.35$ in at least 80\% of usable trajectories. Time error bars show one standard error, the larger of the independent-replica and propagated window errors. A rate comparison requires the two equilibrium means to agree within five combined standard errors. For independent rates $R_a,R_b$ with standard errors $s_a,s_b$, the individual approximate 95\% interval uses the log-ratio standard error $[(s_a/R_a)^2+(s_b/R_b)^2]^{1/2}$.

We fit $\log\tint=\log A+z\log L$ by inverse-relative-variance weighting, separately for each method, observable, and order/window regime of cumulative matching. A fit requires at least four sizes spanning a factor of four, $p\ge0.05$, and a 95\% half-width of at most 0.05. Dropping either endpoint or using only the upper four sizes must change the slope by at most 0.10 including uncertainty. We check predictions at larger sizes wherever available. Tables~\ref{tab:structure}, \ref{tab:spinmodes}, and~\ref{tab:clockslopes} flag fits ending at the data boundary and descriptive slopes for unstable series. The statistical intervals are conditional on the chosen window.

\onecolumn
\section{Transport of the active site and the long-wavelength height mode}\label{app:transport}
A trajectory of the active site can cover a large distance while doing little to a particular Fourier amplitude. To separate the two, we vary the stationary current of nearest-neighbor lifting by mixing its generator with a time-reversed control,
\begin{equation}
 Q_\rho=\tfrac12(1+\rho)Q+\tfrac12(1-\rho)Q^\dagger,\qquad 0\leq\rho\leq1,
\end{equation}
where $Q^\dagger$ is the stationary time reversal of the lifted generator $Q$. $Q$ and $Q^\dagger$ have the same escape rates and stationary distribution but opposite stationary probability currents, so $Q_\rho$ keeps the escape rates and multiplies every stationary current by $\rho$. The reversible control has $\rho=0$, and the proposed sampler has $\rho=1$. Summing over directions cancels the instantaneous marginal current on physical links.

\begin{figure}[!htbp]
\centering\includegraphics[width=\linewidth]{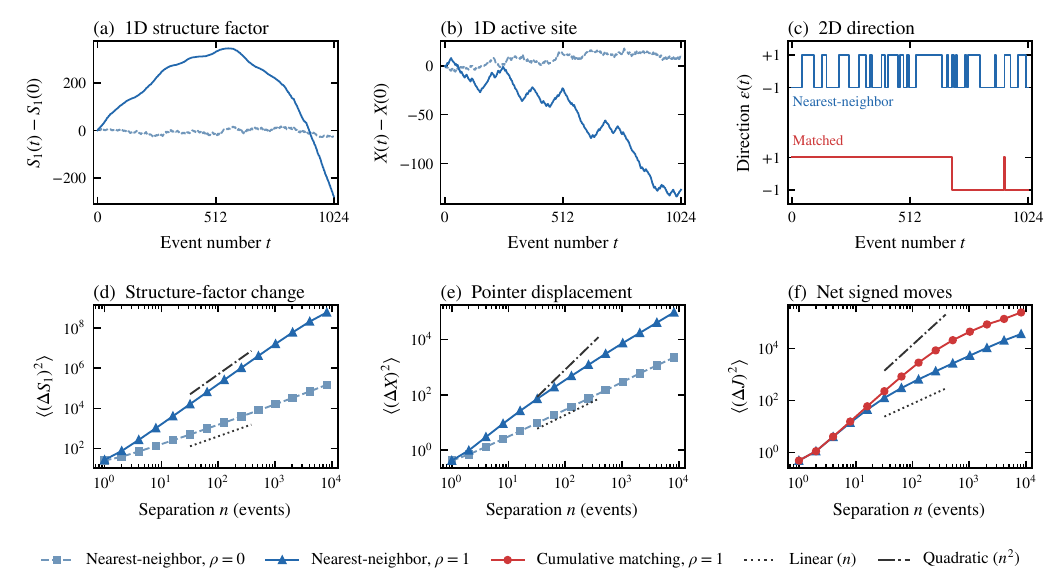}
\caption[Event trajectories and mean-square displacements]{Event trajectories in the discrete Gaussian model and mean-square displacements over $n$ events. (a,b) Structure factor $S_1$ and unwrapped coordinate $X$ of the active site on the 1D chain ($K=0.5$, $L=1024$), for the reversible control ($\rho=0$, dashed) and the proposed sampler ($\rho=1$, solid). (c) Direction $\epsilon$ on the 2D lattice ($K=0.6645$, $L=64$) for nearest-neighbor lifting (upper track) and cumulative matching (lower track), both at $\rho=1$. (d,e) At $\rho=0$ the slopes in (d) and (e) stay near 1. At $\rho=1$ the slope for $S_1$ stays near 2 at short and intermediate separations, while the slope for $X$ decreases toward 1. (f) At $\rho=1$ in 2D, the slope for the net signed move count $J$ falls from about 2 to below 1 with cumulative matching (0.81 over 1024--4096 events).}\label{fig:transport}
\end{figure}

\setlength{\intextsep}{6pt plus 1pt minus 1pt}
\setlength{\textfloatsep}{8pt plus 1pt minus 1pt}
\setlength{\floatsep}{6pt plus 1pt minus 1pt}
\setcounter{topnumber}{4}
\setcounter{totalnumber}{5}

Figure~\ref{fig:transport}(a--c) shows 1024 events from one trajectory per setting. Panels (d--f) average squared displacements over overlapping origins in eight independent trajectories of 131\,072 events each. The 2D matched trace uses $W=256$, mean observation interval 1.2766, and 313 burn-in observations. Each slope uses three successive powers of two over lags 8--32, 64--256, or 1024--4096. We compare slopes by paired jackknife over trajectories and two-sided $t$ tests (seven degrees of freedom), with Holm's correction over 18 comparisons at familywise level 0.05 \cite{holm}.

In 1D at $\rho=0$, the slopes for both $S_1$ and the unwrapped coordinate $X$ stay near one, and only the short-to-intermediate change for $S_1$ is significant. At $\rho=1$, the short and intermediate $S_1$ slopes approach two, and the $X$ slope falls with lag; all three window pairs differ significantly for each full-flow curve. Both processes keep the same direction throughout. In square-lattice DG at $K=0.6645$, $L=64$, the slopes of the net signed move count $J$ are 1.587, 1.067, and 0.965 with nearest-neighbor lifting, and 1.940, 1.667, and 0.813 with matching.

A simple bound relates these displacements to the relaxation time. For $\av{S_1}\propto L^2$ an unnormalized Fourier amplitude has rms magnitude $(N\av{S_1})^{1/2}\sim L^{1+d/2}$, and each unit height move changes that amplitude by at most one. A typical excursion therefore requires at least of order $L^{1-d/2}$ sweeps, or $L^{1/2}$ in 1D. The signed correlation integral can also receive negative contributions, so this is not a strict bound on $\tint$; the fitted 1D exponent $0.507\pm0.016$ nevertheless lies close to it. Persistent handoffs redistribute density in particle chains \cite{kapfer}, the downhill pointer in a harmonic chain relates to self-repelling motion \cite{maggs,localtime,amit,toth}, and bricklayer growth processes link deposition to growth rules or walker crossings \cite{brickgrowth,deposition,bricklayer}. In the DG event chain, by contrast, a height increment and a handoff are separate events.

\section{Size dependence of the long-wavelength observables}\label{app:modes}
Tables~\ref{tab:structure} and~\ref{tab:spinmodes} give the method-specific size windows for $S_1$. Figures~\ref{fig:sizes} and~\ref{fig:spins} compare the measured times at common $L$.

\begin{table}[!htp]
\centering
\caption{Slopes for $\tint(S_1)$ in Fig.~\ref{fig:sizes}, with approximate 95\% half-widths and fit ranges. Plain entries pass checks on larger sizes. An asterisk marks fits whose endpoint is the largest admitted size. A dagger gives the descriptive slope of the largest four sizes when no window passes the criteria.}\label{tab:structure}
\smallskip
\footnotesize
\setlength{\tabcolsep}{3pt}
\begin{tabular*}{0.99\linewidth}{@{\extracolsep{\fill}}llcccccc@{}}
\hline
Model & $K$ & Metro. & Heat bath & NN lifting & NN + matching & IMGS & Suwa--Todo \\
\hline
1D DG & 0.5 & \shortstack{$1.98\pm0.03$\\[-1pt]{\scriptsize 32--256}} & \shortstack{$2.01\pm0.02$\\[-1pt]{\scriptsize 32--512}} & \shortstack{$0.51\pm0.02$\\[-1pt]{\scriptsize 32--256}} & \shortstack{$0.50\pm0.02$\\[-1pt]{\scriptsize 32--256}} & \shortstack{$2.00\pm0.03$\\[-1pt]{\scriptsize 32--256}} & \shortstack{$1.99\pm0.01$\\[-1pt]{\scriptsize 32--512}} \\[1pt]
2D DG & 0.1 & \shortstack{$1.97\pm0.02$\\[-1pt]{\scriptsize 8--64}} & \shortstack{$2.00\pm0.02$\\[-1pt]{\scriptsize 16--128}} & \shortstack{$0.13\pm0.02$\\[-1pt]{\scriptsize 32--256}} & \shortstack{$0.13\pm0.03$\\[-1pt]{\scriptsize 32--256}} & \shortstack{$2.00\pm0.02^{*}$\\[-1pt]{\scriptsize 8--128}} & \shortstack{$1.98\pm0.01$\\[-1pt]{\scriptsize 8--64}} \\[1pt]
2D DG & 0.3 & \shortstack{$1.97\pm0.02$\\[-1pt]{\scriptsize 8--64}} & \shortstack{$1.99\pm0.01$\\[-1pt]{\scriptsize 16--128}} & \shortstack{$1.35\pm0.27^{\dagger}$\\[-1pt]{\scriptsize 64--512}} & \shortstack{$0.18\pm0.02$\\[-1pt]{\scriptsize 16--256}} & \shortstack{$1.99\pm0.02$\\[-1pt]{\scriptsize 8--128}} & \shortstack{$1.99\pm0.02$\\[-1pt]{\scriptsize 16--128}} \\[1pt]
2D DG & 0.6645 & \shortstack{$1.95\pm0.03$\\[-1pt]{\scriptsize 16--128}} & \shortstack{$1.99\pm0.02^{*}$\\[-1pt]{\scriptsize 32--256}} & \shortstack{$1.89\pm0.03$\\[-1pt]{\scriptsize 32--256}} & \shortstack{$1.90\pm0.06^{\dagger}$\\[-1pt]{\scriptsize 64--512}} & \shortstack{$2.05\pm0.03$\\[-1pt]{\scriptsize 8--64}} & \shortstack{$1.96\pm0.01^{*}$\\[-1pt]{\scriptsize 16--256}} \\[1pt]
3D DG & 0.2 & \shortstack{$1.97\pm0.03^{\dagger}$\\[-1pt]{\scriptsize 8--64}} & \shortstack{$1.96\pm0.02^{\dagger}$\\[-1pt]{\scriptsize 8--64}} & \shortstack{$0.67\pm0.16^{\dagger}$\\[-1pt]{\scriptsize 16--128}} & \shortstack{$-0.10\pm0.05^{\dagger}$\\[-1pt]{\scriptsize 16--128}} & \shortstack{$1.95\pm0.04^{\dagger}$\\[-1pt]{\scriptsize 8--64}} & \shortstack{$1.96\pm0.03^{*}$\\[-1pt]{\scriptsize 8--64}} \\[1pt]
3D DG & 0.25 & \shortstack{$1.66\pm0.20^{\dagger}$\\[-1pt]{\scriptsize 8--64}} & \shortstack{$1.60\pm0.20^{\dagger}$\\[-1pt]{\scriptsize 8--64}} & \shortstack{$0.52\pm0.04^{\dagger}$\\[-1pt]{\scriptsize 16--128}} & \shortstack{$-0.62\pm0.21^{\dagger}$\\[-1pt]{\scriptsize 16--128}} & \shortstack{$1.64\pm0.12^{\dagger}$\\[-1pt]{\scriptsize 8--64}} & \shortstack{$1.64\pm0.18^{\dagger}$\\[-1pt]{\scriptsize 8--64}} \\[1pt]
2D clock & 1.25 & \shortstack{$1.93\pm0.10^{\dagger}$\\[-1pt]{\scriptsize 8--64}} & \shortstack{$1.99\pm0.10^{\dagger}$\\[-1pt]{\scriptsize 16--128}} & \shortstack{$1.92\pm0.05^{\dagger}$\\[-1pt]{\scriptsize 32--256}} & \shortstack{$1.60\pm0.19^{\dagger}$\\[-1pt]{\scriptsize 32--256}} & \shortstack{$2.03\pm0.04$\\[-1pt]{\scriptsize 8--64}} & \shortstack{$1.90\pm0.08^{\dagger}$\\[-1pt]{\scriptsize 16--128}} \\[1pt]
3D clock & 0.4542 & \shortstack{$1.87\pm0.08^{\dagger}$\\[-1pt]{\scriptsize 4--32}} & \shortstack{$1.88\pm0.09^{\dagger}$\\[-1pt]{\scriptsize 4--32}} & \shortstack{$1.47\pm0.12^{\dagger}$\\[-1pt]{\scriptsize 8--64}} & \shortstack{$1.41\pm0.07^{\dagger}$\\[-1pt]{\scriptsize 8--64}} & \shortstack{$1.88\pm0.05^{\dagger}$\\[-1pt]{\scriptsize 4--32}} & \shortstack{$1.88\pm0.09^{\dagger}$\\[-1pt]{\scriptsize 4--32}} \\[1pt]
\hline
\end{tabular*}
\par\bigskip
\normalsize
\caption{Finite-size slopes for $\tint(S_1)$ of the Ising and Potts models in Fig.~\ref{fig:spins}. Intervals and flags follow Table~\ref{tab:structure}. The displayed cumulative series have $W=64$ throughout each fit, with linear, Hilbert, and serpentine site order in 1D, 2D, and 3D, respectively. The 1D temperatures are fixed. The 3D Potts model is at first-order coexistence.}\label{tab:spinmodes}
\smallskip
\footnotesize
\setlength{\tabcolsep}{3pt}
\begin{tabular*}{0.99\linewidth}{@{\extracolsep{\fill}}llcccccc@{}}
\hline
Model & $K$ & Metro. & Heat bath & NN lifting & NN + matching & IMGS & Suwa--Todo \\
\hline
1D Ising & 2.7739 & \shortstack{$0.11\pm0.04^{\dagger}$\\[-1pt]{\scriptsize 128--1024}} & \shortstack{$0.10\pm0.06^{\dagger}$\\[-1pt]{\scriptsize 128--1024}} & \shortstack{$-0.10\pm0.02^{*}$\\[-1pt]{\scriptsize 128--1024}} & \shortstack{$-0.12\pm0.02^{*}$\\[-1pt]{\scriptsize 128--1024}} & \shortstack{$0.43\pm0.16^{\dagger}$\\[-1pt]{\scriptsize 128--1024}} & \shortstack{$0.10\pm0.06^{\dagger}$\\[-1pt]{\scriptsize 128--1024}} \\[1pt]
2D Ising & 0.88137 & \shortstack{$2.21\pm0.09^{\dagger}$\\[-1pt]{\scriptsize 48--128}} & \shortstack{$2.13\pm0.04$\\[-1pt]{\scriptsize 6--32}} & \shortstack{$2.18\pm0.03^{*}$\\[-1pt]{\scriptsize 12--128}} & \shortstack{$2.14\pm0.04^{*}$\\[-1pt]{\scriptsize 16--128}} & \shortstack{$2.51\pm0.16^{\dagger}$\\[-1pt]{\scriptsize 48--128}} & \shortstack{$2.13\pm0.04^{*}$\\[-1pt]{\scriptsize 8--96}} \\[1pt]
3D Ising & 0.44331 & \shortstack{$1.95\pm0.03^{*}$\\[-1pt]{\scriptsize 6--32}} & \shortstack{$1.89\pm0.03$\\[-1pt]{\scriptsize 6--24}} & \shortstack{$1.90\pm0.09^{\dagger}$\\[-1pt]{\scriptsize 16--32}} & \shortstack{$1.69\pm0.11^{\dagger}$\\[-1pt]{\scriptsize 6--24}} & \shortstack{$2.19\pm0.04^{\dagger}$\\[-1pt]{\scriptsize 16--32}} & \shortstack{$1.95\pm0.14^{\dagger}$\\[-1pt]{\scriptsize 16--32}} \\[1pt]
1D Potts & 3.1583 & \shortstack{$0.09\pm0.07^{\dagger}$\\[-1pt]{\scriptsize 128--1024}} & \shortstack{$0.09\pm0.07^{\dagger}$\\[-1pt]{\scriptsize 128--1024}} & \shortstack{$0.00\pm0.02$\\[-1pt]{\scriptsize 64--512}} & \shortstack{$-0.04\pm0.02^{\dagger}$\\[-1pt]{\scriptsize 128--1024}} & \shortstack{$0.34\pm0.14^{\dagger}$\\[-1pt]{\scriptsize 128--1024}} & \shortstack{$0.09\pm0.07^{\dagger}$\\[-1pt]{\scriptsize 128--1024}} \\[1pt]
2D Potts & 1.0051 & \shortstack{$2.16\pm0.03^{*}$\\[-1pt]{\scriptsize 8--96}} & \shortstack{$2.14\pm0.04^{*}$\\[-1pt]{\scriptsize 12--96}} & \shortstack{$2.14\pm0.11^{\dagger}$\\[-1pt]{\scriptsize 48--128}} & \shortstack{$2.15\pm0.07^{\dagger}$\\[-1pt]{\scriptsize 16--128}} & \shortstack{$2.36\pm0.07^{\dagger}$\\[-1pt]{\scriptsize 32--96}} & \shortstack{$2.07\pm0.13^{\dagger}$\\[-1pt]{\scriptsize 48--128}} \\[1pt]
3D Potts & 0.55056 & \shortstack{$3.19\pm0.22^{\dagger}$\\[-1pt]{\scriptsize 16--32}} & \shortstack{$3.07\pm0.23^{\dagger}$\\[-1pt]{\scriptsize 16--32}} & \shortstack{$3.05\pm0.28^{\dagger}$\\[-1pt]{\scriptsize 16--32}} & \shortstack{$2.52\pm0.15^{\dagger}$\\[-1pt]{\scriptsize 6--24}} & \shortstack{$3.10\pm0.33^{\dagger}$\\[-1pt]{\scriptsize 16--32}} & \shortstack{$2.94\pm0.33^{\dagger}$\\[-1pt]{\scriptsize 16--32}} \\[1pt]
\hline
\end{tabular*}
\end{table}

\section{Energy and magnetic relaxation in the six-state clock model}
The energy per bond, the fluctuations at wavelength $L$, and the uniform magnetization amplitude can decorrelate on different time scales. In two dimensions matching speeds up both magnetic quantities at the larger sizes. In three dimensions the uniform amplitude remains slower than $S_1$. Table~\ref{tab:clockslopes} records the separate fit ranges and uncertainties for each mode.

\begin{figure}[!htp]
\centering\includegraphics[width=0.88\linewidth]{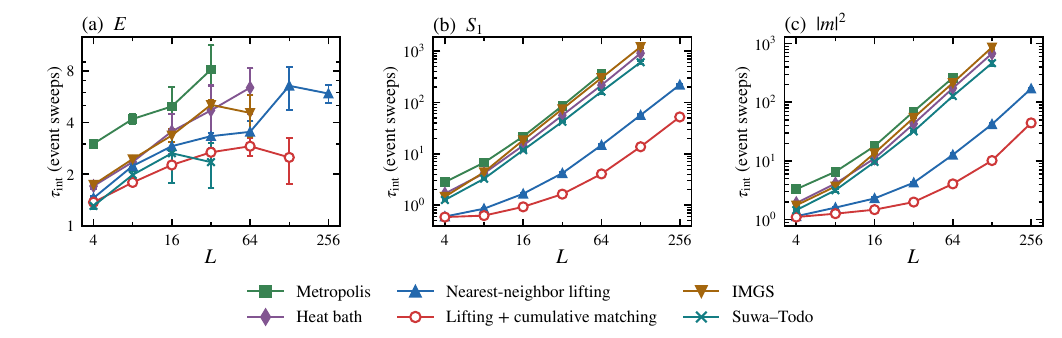}
\caption{Integrated autocorrelation time versus lattice size in the square-lattice six-state clock model at $K=1.25$ for (a) the energy per bond $E$, (b) the longest-wavelength structure factor $S_1$, and (c) the squared magnetization $|m|^2$. Units, symbols, and error bars follow Fig.~\ref{fig:sizes}. Table~\ref{tab:clockslopes} gives the size exponents for (b) and (c).}\label{fig:clockplane}
\end{figure}

\begin{figure}[!htp]
\centering\includegraphics[width=0.88\linewidth]{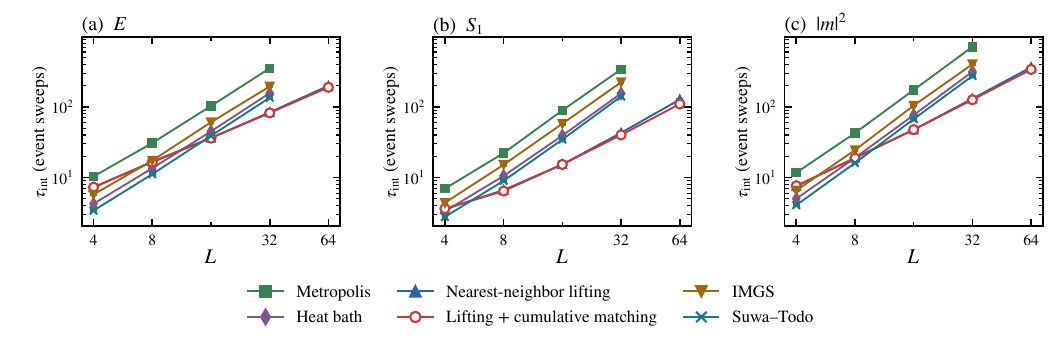}
\caption{Integrated autocorrelation time in the cubic-lattice six-state clock model at $K=0.4542$ for (a) the energy per bond $E$, (b) the longest-wavelength structure factor $S_1$, and (c) the squared magnetization $|m|^2$. All panels share the same vertical scale in event sweeps. Table~\ref{tab:clockslopes} gives the fitted exponent and size range for each curve.}\label{fig:clock}
\end{figure}

\begin{table}[!htp]
\centering
\caption{Effective size exponents for $S_1$ and $|m|^2$ in the 2D clock model at $K=1.25$ [Fig.~\ref{fig:clockplane}(b,c)] and for $E$, $S_1$, and $|m|^2$ in the 3D clock model at $K=0.4542$ [Fig.~\ref{fig:clock}]. Each fitted range appears under its slope. The intervals and symbols follow Table~\ref{tab:structure}.}\label{tab:clockslopes}
\smallskip
\footnotesize
\setlength{\tabcolsep}{3pt}
\begin{tabular*}{0.99\linewidth}{@{\extracolsep{\fill}}llcccccc@{}}
\hline
Model & Obs. & Metro. & Heat bath & NN lifting & NN + matching & IMGS & Suwa--Todo \\
\hline
2D clock & $S_1$ & \shortstack{$1.93\pm0.10^{\dagger}$\\[-1pt]{\scriptsize 8--64}} & \shortstack{$1.99\pm0.10^{\dagger}$\\[-1pt]{\scriptsize 16--128}} & \shortstack{$1.92\pm0.05^{\dagger}$\\[-1pt]{\scriptsize 32--256}} & \shortstack{$1.60\pm0.19^{\dagger}$\\[-1pt]{\scriptsize 32--256}} & \shortstack{$2.03\pm0.04$\\[-1pt]{\scriptsize 8--64}} & \shortstack{$1.90\pm0.08^{\dagger}$\\[-1pt]{\scriptsize 16--128}} \\[1pt]
2D clock & $|m|^2$ & \shortstack{$1.80\pm0.12^{\dagger}$\\[-1pt]{\scriptsize 8--64}} & \shortstack{$1.99\pm0.12^{\dagger}$\\[-1pt]{\scriptsize 16--128}} & \shortstack{$1.80\pm0.12^{\dagger}$\\[-1pt]{\scriptsize 32--256}} & \shortstack{$1.31\pm0.31^{\dagger}$\\[-1pt]{\scriptsize 32--256}} & \shortstack{$1.97\pm0.04^{*}$\\[-1pt]{\scriptsize 8--128}} & \shortstack{$1.88\pm0.12^{\dagger}$\\[-1pt]{\scriptsize 16--128}} \\[1pt]
3D clock & $E$ & \shortstack{$1.69\pm0.06^{\dagger}$\\[-1pt]{\scriptsize 4--32}} & \shortstack{$1.75\pm0.04^{\dagger}$\\[-1pt]{\scriptsize 4--32}} & \shortstack{$1.21\pm0.05^{\dagger}$\\[-1pt]{\scriptsize 8--64}} & \shortstack{$1.19\pm0.02^{*}$\\[-1pt]{\scriptsize 8--64}} & \shortstack{$1.70\pm0.05^{\dagger}$\\[-1pt]{\scriptsize 4--32}} & \shortstack{$1.77\pm0.03^{*}$\\[-1pt]{\scriptsize 4--32}} \\[1pt]
3D clock & $S_1$ & \shortstack{$1.87\pm0.08^{\dagger}$\\[-1pt]{\scriptsize 4--32}} & \shortstack{$1.88\pm0.09^{\dagger}$\\[-1pt]{\scriptsize 4--32}} & \shortstack{$1.47\pm0.12^{\dagger}$\\[-1pt]{\scriptsize 8--64}} & \shortstack{$1.41\pm0.07^{\dagger}$\\[-1pt]{\scriptsize 8--64}} & \shortstack{$1.88\pm0.05^{\dagger}$\\[-1pt]{\scriptsize 4--32}} & \shortstack{$1.88\pm0.09^{\dagger}$\\[-1pt]{\scriptsize 4--32}} \\[1pt]
3D clock & $|m|^2$ & \shortstack{$1.97\pm0.05^{\dagger}$\\[-1pt]{\scriptsize 4--32}} & \shortstack{$2.01\pm0.03^{\dagger}$\\[-1pt]{\scriptsize 4--32}} & \shortstack{$1.45\pm0.05^{\dagger}$\\[-1pt]{\scriptsize 8--64}} & \shortstack{$1.41\pm0.03^{\dagger}$\\[-1pt]{\scriptsize 8--64}} & \shortstack{$1.99\pm0.05^{\dagger}$\\[-1pt]{\scriptsize 4--32}} & \shortstack{$2.03\pm0.03^{*}$\\[-1pt]{\scriptsize 4--32}} \\[1pt]
\hline
\end{tabular*}
\end{table}

\setlength{\abovecaptionskip}{4pt}
\section{Production rates by observable}
At the size and recording interval of Fig.~\ref{fig:gain}, the points give rate ratios to Metropolis with individual 95\% intervals for each recorded observable shown.

\begin{figure}[!htp]
\centering\includegraphics[width=0.80\linewidth]{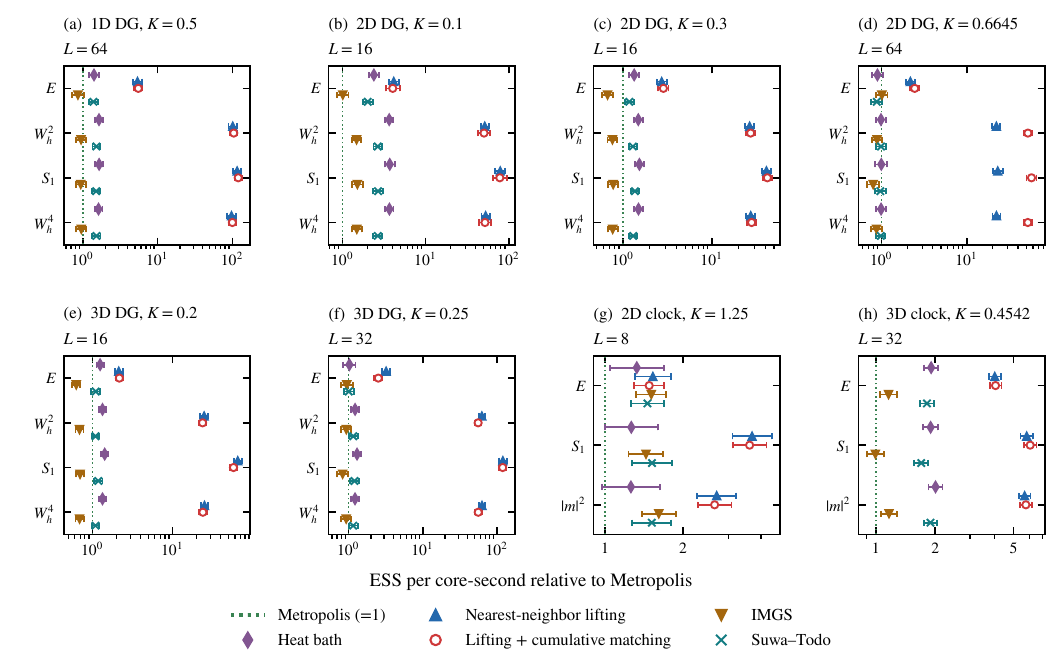}
\caption{Effective samples per production core-second relative to Metropolis (vertical dotted line) for DG and clock. The size is chosen using $S_1$. For DG, $W_h^4=(W_h^2)^2$.}\label{fig:efficiency}
\end{figure}

\begin{figure}[!htp]
\centering\includegraphics[width=0.80\linewidth]{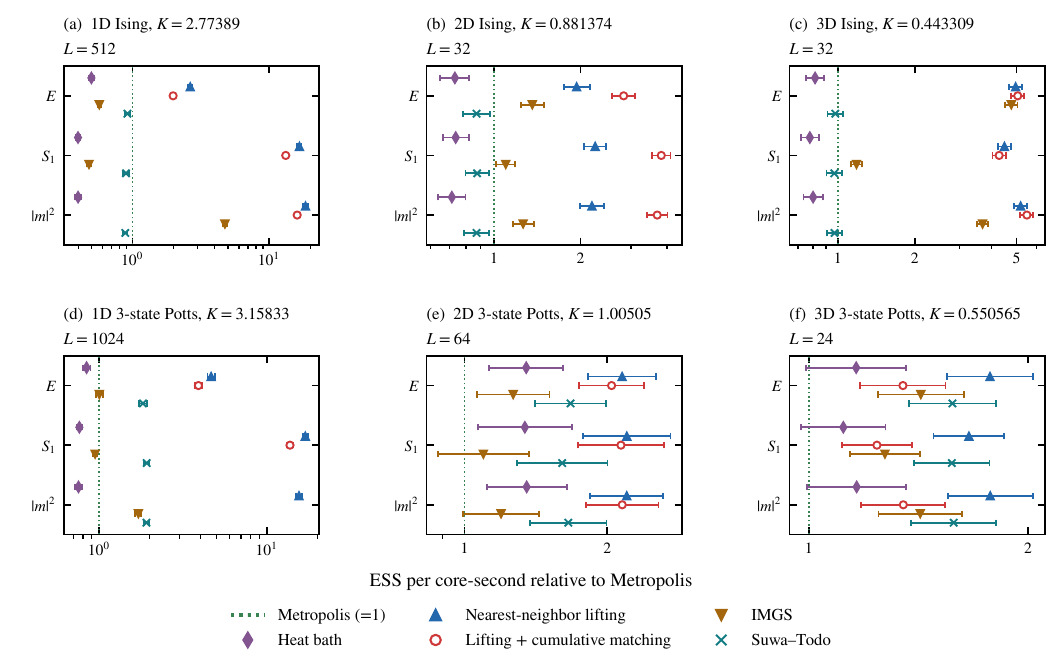}
\caption{Effective samples per production core-second relative to Metropolis for Ising (top) and three-state Potts (bottom). The size is chosen using $S_1$.}\label{fig:spincost}
\end{figure}

\twocolumn


\begin{thebibliography}{99}

\bibitem{metropolis}
N. Metropolis, A. W. Rosenbluth, M. N. Rosenbluth, A. H. Teller, and E. Teller, Equation of state calculations by fast computing machines,
\doi{10.1063/1.1699114}{J. Chem. Phys. \textbf{21}, 1087--1092 (1953)}.

\bibitem{hastings}
W. K. Hastings, Monte Carlo sampling methods using Markov chains and their applications,
\doi{10.1093/biomet/57.1.97}{Biometrika \textbf{57}, 97--109 (1970)}.

\bibitem{sokal}
A. Sokal, Monte Carlo methods in statistical mechanics: Foundations and new algorithms,
in \doi{10.1007/978-1-4899-0319-8_6}{\emph{Functional Integration: Basics and Applications}},
edited by C. DeWitt-Morette, P. Cartier, and A. Folacci (Plenum, New York, 1997), pp.~131--192.

\bibitem{chui}
S. T. Chui and J. D. Weeks, Phase transition in the two-dimensional Coulomb gas, and the interfacial roughening transition,
\doi{10.1103/PhysRevB.14.4978}{Phys. Rev. B \textbf{14}, 4978--4982 (1976)}.

\bibitem{rough}
M. Hasenbusch, M. Marcu, and K. Pinn, High precision renormalization group study of the roughening transition,
\doi{10.1016/0378-4371(94)90536-3}{Physica A \textbf{208}, 124--161 (1994)}.

\bibitem{rough2}
M. Hasenbusch and K. Pinn, Computing the roughening transition of Ising and solid-on-solid models by BCSOS model matching,
\doi{10.1088/0305-4470/30/1/006}{J. Phys. A \textbf{30}, 63--80 (1997)}.

\bibitem{hohenberg}
P. C. Hohenberg and B. I. Halperin, Theory of dynamic critical phenomena,
\doi{10.1103/RevModPhys.49.435}{Rev. Mod. Phys. \textbf{49}, 435--479 (1977)}.

\bibitem{creutz}
M. Creutz, Overrelaxation and Monte Carlo simulation,
\doi{10.1103/PhysRevD.36.515}{Phys. Rev. D \textbf{36}, 515--519 (1987)}.

\bibitem{brown}
F. R. Brown and T. J. Woch, Overrelaxed heat-bath and Metropolis algorithms for accelerating pure gauge Monte Carlo calculations,
\doi{10.1103/PhysRevLett.58.2394}{Phys. Rev. Lett. \textbf{58}, 2394--2396 (1987)}.

\bibitem{chen}
F. Chen, L. Lov\'asz, and I. Pak, Lifting Markov chains to speed up mixing,
in \doi{10.1145/301250.301315}{\emph{Proceedings of the Thirty-First Annual ACM Symposium on Theory of Computing}} (ACM, 1999), pp.~275--281.

\bibitem{diaconis}
P. Diaconis, S. Holmes, and R. M. Neal, Analysis of a nonreversible Markov chain sampler,
\doi{10.1214/aoap/1019487508}{Ann. Appl. Probab. \textbf{10}, 726--752 (2000)}.

\bibitem{gustafson}
P. Gustafson, A guided walk Metropolis algorithm,
\doi{10.1023/A:1008880707168}{Stat. Comput. \textbf{8}, 357--364 (1998)}.

\bibitem{turitsyn}
K. S. Turitsyn, M. Chertkov, and M. Vucelja, Irreversible Monte Carlo algorithms for efficient sampling,
\doi{10.1016/j.physd.2010.10.003}{Physica D \textbf{240}, 410--414 (2011)}.

\bibitem{fernandes}
H. C. M. Fernandes and M. Weigel, Non-reversible Monte Carlo simulations of spin models,
\doi{10.1016/j.cpc.2010.11.017}{Comput. Phys. Commun. \textbf{182}, 1856--1859 (2011)}.

\bibitem{sakaiising}
Y. Sakai and K. Hukushima, Dynamics of one-dimensional Ising model without detailed balance condition,
\doi{10.7566/JPSJ.82.064003}{J. Phys. Soc. Jpn. \textbf{82}, 064003 (2013)}.

\bibitem{curieweiss}
J. Bierkens and G. Roberts, A piecewise deterministic scaling limit of lifted Metropolis--Hastings in the Curie--Weiss model,
\doi{10.1214/16-AAP1217}{Ann. Appl. Probab. \textbf{27}, 846--882 (2017)}.

\bibitem{suwa}
H. Suwa and S. Todo, Markov chain Monte Carlo method without detailed balance,
\doi{10.1103/PhysRevLett.105.120603}{Phys. Rev. Lett. \textbf{105}, 120603 (2010)}.

\bibitem{allocation}
H. Suwa, Geometric allocation approach in Markov chain Monte Carlo, in
\doi{10.1007/978-4-431-54517-0_2}{\emph{Geometrically Constructed Markov Chain Monte Carlo Study of Quantum Spin-phonon Complex Systems}} (Springer, Tokyo, 2014), pp.~11--42.

\bibitem{sakairing}
Y. Sakai and K. Hukushima, Eigenvalue analysis of an irreversible random walk with skew detailed balance conditions,
\doi{10.1103/PhysRevE.93.043318}{Phys. Rev. E \textbf{93}, 043318 (2016)}.

\bibitem{gagnon}
P. Gagnon and F. Maire, Theoretical guarantees for lifted samplers,
\doi{10.1016/j.spa.2026.104937}{Stochastic Process. Appl. \textbf{199}, 104937 (2026)}.

\bibitem{vucelja}
M. Vucelja, Lifting---A nonreversible Markov chain Monte Carlo algorithm,
\doi{10.1119/1.4961596}{Am. J. Phys. \textbf{84}, 958--968 (2016)}.

\bibitem{flowreview}
H. Suwa and S. Todo, Control of probability flow in Markov chain Monte Carlo---Nonreversibility and lifting,
\doi{10.1063/5.0233858}{J. Chem. Phys. \textbf{161}, 174107 (2024)}.

\bibitem{bernard}
E. P. Bernard, W. Krauth, and D. B. Wilson, Event-chain Monte Carlo algorithms for hard-sphere systems,
\doi{10.1103/PhysRevE.80.056704}{Phys. Rev. E \textbf{80}, 056704 (2009)}.

\bibitem{michel}
M. Michel, S. C. Kapfer, and W. Krauth, Generalized event-chain Monte Carlo: Constructing rejection-free global-balance algorithms from infinitesimal steps,
\doi{10.1063/1.4863991}{J. Chem. Phys. \textbf{140}, 054116 (2014)}.

\bibitem{krauthreview}
W. Krauth, Event-chain Monte Carlo: Foundations, applications, and prospects,
\doi{10.3389/fphy.2021.663457}{Front. Phys. \textbf{9}, 663457 (2021)}.

\bibitem{spin}
M. Michel, J. Mayer, and W. Krauth, Event-chain Monte Carlo for classical continuous spin models,
\doi{10.1209/0295-5075/112/20003}{EPL \textbf{112}, 20003 (2015)}.

\bibitem{realfield}
M. Hasenbusch and S. Schaefer, Testing the event-chain algorithm in asymptotically free models,
\doi{10.1103/PhysRevD.98.054502}{Phys. Rev. D \textbf{98}, 054502 (2018)}.

\bibitem{essler}
F. H. L. Essler and W. Krauth, Lifted TASEP: A solvable paradigm for speeding up many-particle Markov chains,
\doi{10.1103/PhysRevX.14.041035}{Phys. Rev. X \textbf{14}, 041035 (2024)}.

\bibitem{localtasep}
B. Massouli\'e, C. Erignoux, C. Toninelli, and W. Krauth, Velocity trapping in the lifted totally asymmetric simple exclusion process and the true self-avoiding random walk,
\doi{10.1103/mqdr-x95j}{Phys. Rev. Lett. \textbf{135}, 127102 (2025)}.

\bibitem{softtasep}
F. H. L. Essler, J. Gipouloux, and W. Krauth, Lifted TASEP: Long-time dynamics, generalizations, and continuum limit,
\doi{10.21468/SciPostPhysCore.8.4.063}{SciPost Phys. Core \textbf{8}, 063 (2025)}.

\bibitem{harland}
J. Harland, M. Michel, T. A. Kampmann, and J. Kierfeld, Event-chain Monte Carlo algorithms for three- and many-particle interactions,
\doi{10.1209/0295-5075/117/30001}{EPL \textbf{117}, 30001 (2017)}.

\bibitem{faizi}
F. Faizi, G. Deligiannidis, and E. Rosta, Efficient irreversible Monte Carlo samplers,
\doi{10.1021/acs.jctc.9b01135}{J. Chem. Theory Comput. \textbf{16}, 2124--2138 (2020)}.

\bibitem{power}
S. Power and J. Vorstrup Goldman, Accelerated sampling on discrete spaces with non-reversible Markov processes,
\href{https://arxiv.org/abs/1912.04681}{arXiv:1912.04681 [stat.CO] (2019)}.

\bibitem{zanella}
G. Zanella, Informed proposals for local MCMC in discrete spaces,
\doi{10.1080/01621459.2019.1585255}{J. Am. Stat. Assoc. \textbf{115}, 852--865 (2020)}.

\bibitem{grathwohl}
W. Grathwohl, K. Swersky, M. Hashemi, D. Duvenaud, and C. Maddison, Oops I took a gradient: Scalable sampling for discrete distributions,
in \href{https://proceedings.mlr.press/v139/grathwohl21a.html}{\emph{Proceedings of the 38th International Conference on Machine Learning}},
Proc. Mach. Learn. Res. \textbf{139}, 3831--3841 (2021).

\bibitem{kinetic}
P. Monmarch\'e, Kinetic walks for sampling,
\doi{10.30757/ALEA.v17-20}{ALEA Lat. Am. J. Probab. Math. Stat. \textbf{17}, 491--530 (2020)}.

\bibitem{hamiltonian}
R. Barboni, S. Grazzi, and G. Zanella, Hamiltonian dynamics for sampling on discrete spaces,
\href{https://arxiv.org/abs/2608.17961}{arXiv:2608.17961 [math.ST] (2026)}.

\bibitem{dhams}
Y. Zhou and Z. Tan, Preconditioned discrete-HAMS: A second-order irreversible discrete sampler,
\doi{10.1007/s11222-026-10907-8}{Stat. Comput. \textbf{36}, 159 (2026)}.

\bibitem{swendsen}
R. H. Swendsen and J.-S. Wang, Nonuniversal critical dynamics in Monte Carlo simulations,
\doi{10.1103/PhysRevLett.58.86}{Phys. Rev. Lett. \textbf{58}, 86--88 (1987)}.

\bibitem{wolffcluster}
U. Wolff, Collective Monte Carlo updating for spin systems,
\doi{10.1103/PhysRevLett.62.361}{Phys. Rev. Lett. \textbf{62}, 361--364 (1989)}.

\bibitem{evertz}
H. G. Evertz, M. Hasenbusch, M. Marcu, K. Pinn, and S. Solomon, Stochastic cluster algorithms for discrete Gaussian (SOS) models,
\doi{10.1016/0370-2693(91)90418-P}{Phys. Lett. B \textbf{254}, 185--191 (1991)}.

\bibitem{classicalworm}
N. Prokof'ev and B. Svistunov, Worm algorithms for classical statistical models,
\doi{10.1103/PhysRevLett.87.160601}{Phys. Rev. Lett. \textbf{87}, 160601 (2001)}.

\bibitem{leeyang}
T. D. Lee and C. N. Yang, Statistical theory of equations of state and phase transitions. II. Lattice gas and Ising model,
\doi{10.1103/PhysRev.87.410}{Phys. Rev. \textbf{87}, 410--419 (1952)}.

\bibitem{steinoperator}
C. Ley, G. Reinert, and Y. Swan, Stein's method for comparison of univariate distributions,
\doi{10.1214/16-PS278}{Probab. Surv. \textbf{14}, 1--52 (2017)}.

\bibitem{stein}
M. Ernst, G. Reinert, and Y. Swan, First-order covariance inequalities via Stein's method,
\doi{10.3150/19-BEJ1182}{Bernoulli \textbf{26}, 2051--2081 (2020)}.

\bibitem{villain}
J. Villain, Theory of one- and two-dimensional magnets with an easy magnetization plane. II. The planar, classical, two-dimensional magnet,
\doi{10.1051/jphys:01975003606058100}{J. Phys. (Paris) \textbf{36}, 581--590 (1975)}.

\bibitem{hilbert}
D. Hilbert, Ueber die stetige Abbildung einer Linie auf ein Fl\"achenst\"uck,
\doi{10.1007/BF01199431}{Math. Ann. \textbf{38}, 459--460 (1891)}.

\bibitem{gopfert}
M. G\"opfert and G. Mack, Proof of confinement of static quarks in 3-dimensional U(1) lattice gauge theory for all values of the coupling constant,
\doi{10.1007/BF01961240}{Commun. Math. Phys. \textbf{82}, 545--606 (1982)}.

\bibitem{jose}
J. V. Jos\'e, L. P. Kadanoff, S. Kirkpatrick, and D. R. Nelson, Renormalization, vortices, and symmetry-breaking perturbations in the two-dimensional planar model,
\doi{10.1103/PhysRevB.16.1217}{Phys. Rev. B \textbf{16}, 1217--1241 (1977)}.

\bibitem{tomita}
Y. Tomita and Y. Okabe, Probability-changing cluster algorithm for two-dimensional XY and clock models,
\doi{10.1103/PhysRevB.65.184405}{Phys. Rev. B \textbf{65}, 184405 (2002)}.

\bibitem{zou}
X. Zou, S. Liu, and W. Guo, Surface critical properties of the three-dimensional clock model,
\doi{10.1103/PhysRevB.106.064420}{Phys. Rev. B \textbf{106}, 064420 (2022)}.

\bibitem{shao}
H. Shao, W. Guo, and A. W. Sandvik, Monte Carlo renormalization flows in the space of relevant and irrelevant operators: Application to three-dimensional clock models,
\doi{10.1103/PhysRevLett.124.080602}{Phys. Rev. Lett. \textbf{124}, 080602 (2020)}.

\bibitem{protonation}
U. Khaniya, J. Mao, R. J. Wei, and M. R. Gunner, Characterizing protein protonation microstates using Monte Carlo sampling,
\doi{10.1021/acs.jpcb.2c00139}{J. Phys. Chem. B \textbf{126}, 2476--2485 (2022)}.

\bibitem{spinphonon}
H. Suwa, \doi{10.1007/978-4-431-54517-0}{\emph{Geometrically Constructed Markov Chain Monte Carlo Study of Quantum Spin-phonon Complex Systems}}, Springer Theses (Springer, Tokyo, 2014).

\bibitem{tnmc}
S. Goto, H. Suwa, and S. Todo, Tensor-network Monte Carlo approach based on time-evolving block decimation,
\href{https://arxiv.org/abs/2609.12466}{arXiv:2609.12466 [cond-mat.stat-mech] (2026)}.

\bibitem{walker}
A. J. Walker, An efficient method for generating discrete random variables with general distributions,
\doi{10.1145/355744.355749}{ACM Trans. Math. Softw. \textbf{3}, 253--256 (1977)}.

\bibitem{allocationproceedings}
S. Todo and H. Suwa, Geometric allocation approaches in Markov chain Monte Carlo,
\doi{10.1088/1742-6596/473/1/012013}{J. Phys. Conf. Ser. \textbf{473}, 012013 (2013)}.

\bibitem{walkgrowth}
H. Hu, X. Chen, and Y. Deng, Irreversible Markov chain Monte Carlo algorithm for self-avoiding walk,
\doi{10.1007/s11467-016-0646-6}{Front. Phys. \textbf{12}, 120503 (2017)}.

\bibitem{isingworm}
E. M. El\c{c}i, J. Grimm, L. Ding, A. Nasrawi, T. M. Garoni, and Y. Deng, Lifted worm algorithm for the Ising model,
\doi{10.1103/PhysRevE.97.042126}{Phys. Rev. E \textbf{97}, 042126 (2018)}.

\bibitem{worm}
H. Suwa, Geometric allocation approach to accelerating directed worm algorithm,
\doi{10.1103/PhysRevE.103.013308}{Phys. Rev. E \textbf{103}, 013308 (2021)}.

\bibitem{liftedworm}
H. Suwa, Lifted directed-worm algorithm,
\doi{10.1103/PhysRevE.106.055306}{Phys. Rev. E \textbf{106}, 055306 (2022)}.

\bibitem{rejection}
H. Suwa, Reducing rejection exponentially improves Markov chain Monte Carlo sampling,
\doi{10.1016/j.physa.2023.129368}{Physica A \textbf{633}, 129368 (2024)}.

\bibitem{wolff}
U. Wolff, Monte Carlo errors with less errors,
\doi{10.1016/S0010-4655(03)00467-3}{Comput. Phys. Commun. \textbf{156}, 143--153 (2004)}.

\bibitem{holm}
S. Holm, A simple sequentially rejective multiple test procedure,
\href{https://www.jstor.org/stable/4615733}{Scand. J. Stat. \textbf{6}, 65--70 (1979)}.

\bibitem{kapfer}
S. C. Kapfer and W. Krauth, Irreversible local Markov chains with rapid convergence towards equilibrium,
\doi{10.1103/PhysRevLett.119.240603}{Phys. Rev. Lett. \textbf{119}, 240603 (2017)}.

\bibitem{maggs}
A. C. Maggs, Event-chain Monte Carlo and the true self-avoiding walk,
\doi{10.1103/PhysRevE.111.054104}{Phys. Rev. E \textbf{111}, 054104 (2025)}.

\bibitem{localtime}
A. Eberle and F. L\"orler, Convergence rates of self-repellent random walks, their local time and Event Chain Monte Carlo,
\href{https://arxiv.org/abs/2511.23453}{arXiv:2511.23453 [math.PR] (2025)}.

\bibitem{amit}
D. J. Amit, G. Parisi, and L. Peliti, Asymptotic behavior of the ``true'' self-avoiding walk,
\doi{10.1103/PhysRevB.27.1635}{Phys. Rev. B \textbf{27}, 1635--1645 (1983)}.

\bibitem{toth}
B. T\'oth and W. Werner, The true self-repelling motion,
\doi{10.1007/s004400050172}{Probab. Theory Relat. Fields \textbf{111}, 375--452 (1998)}.

\bibitem{brickgrowth}
M. Bal\'azs, Microscopic shape of shocks in a domain growth model,
\doi{10.1023/A:1012271624597}{J. Stat. Phys. \textbf{105}, 511--524 (2001)}.

\bibitem{deposition}
B. T\'oth and W. Werner, Hydrodynamic equation for a deposition model, in
\doi{10.1007/978-1-4612-0063-5_9}{\emph{In and Out of Equilibrium: Probability with a Physics Flavor}},
edited by V. Sidoravicius, Progress in Probability \textbf{51} (Birkh\"auser, 2002), pp.~227--248.

\bibitem{bricklayer}
A. C. Maggs, Dynamics of a bricklayer model: Multi-walker realizations of true self-avoiding motion,
\doi{10.1007/s10955-026-03594-2}{J. Stat. Phys. \textbf{193}, 39 (2026)}.

\end{thebibliography}
\end{document}